\documentclass[preprint,showpacs,preprintnumbers,amsmath,amssymb,nofootinbib]{revtex4}

\usepackage{amsthm}

{Corollary}[section]

\usepackage{etex}
\usepackage{amssymb,amsthm,amscd,amsbsy,array}
\usepackage{bm}
\usepackage{amsmath,dsfont}
\usepackage{soul} 
\usepackage{graphics,graphicx,xcolor}
\usepackage[colorlinks=true, pdfstartview=FitV, linkcolor=blue, citecolor=blue, urlcolor=blue]{hyperref} %
\usepackage{algorithm}
\usepackage{algorithmic}
\usepackage{longtable}
\usepackage{epstopdf}
\usepackage{subfigure}
\usepackage{float}
\usepackage[T1]{fontenc}
\usepackage[english]{babel}
\usepackage{graphicx}
\usepackage{float}
\usepackage{tikz}
\usepackage{mathrsfs}
\usepackage[scr=rsfs,cal=boondox]{mathalfa}

\newcommand{\red}{\textcolor{red}}
\newcommand{\blue}{\textcolor{blue}}

\newcommand{\orange}{\textcolor{orange}}
\newcommand{\gb}{\colorbox{green}}

\newcommand{\magenta}{\textcolor{magenta}}
\newenvironment{redtext}{\color{red}}{\ignorespacesafterend}
\newenvironment{bluetext}{\color{blue}}{\ignorespacesafterend}

\newenvironment{magentatext}{\color{magenta}}{\ignorespacesafterend}
\newenvironment{orangetext}{\color{orange}}{\ignorespacesafterend}
\newenvironment{cyantext}{\color{cyan}}{\ignorespacesafterend}
\newcommand{\bblue}{\begin{bluetext}}
\newcommand{\eblue}{\end{bluetext}}
\newcommand{\bred}{\begin{redtext}}
\newcommand{\ered}{\end{redtext}}
\newcommand{\bmagenta}{\begin{magentatext}}
\newcommand{\emagenta}{\end{magentatext}}
\newcommand{\borange}{\begin{orangetext}}
\newcommand{\eorange}{\end{orangetext}}
\newcommand{\bcyan}{\begin{cyantext}}
\newcommand{\ecyan}{\end{cyantext}}

\numberwithin{equation}{section}

\let\ssection=\section
\renewcommand{\section}{\setcounter{equation}{0}\ssection}

\newcommand{\cA}{{\mathcal{A}}}

\newcommand{\PT}{{P\"oschl{\strut}-Teller\;}}
\newcommand{\GW}{{gravitational wave\;}}
\newcommand{\GWs}{{gravitational waves\;}}

\newcommand{\cL}{{\mathcal{L}}}

\newcommand{\bX}{{\bf X}}

\def\smallover#1/#2{\hbox{$\textstyle\frac{#1}{#2}$}} %

\def\besub{\begin{subequations}}
\def\esub{\end{subequations}}
\def\benu{\begin{enumerate}}
\def\eenu{\end{enumerate}}
\def\beq{\begin{equation}}
\def\eeq{\end{equation}}
\def\beqa{\begin{eqnarray}}
\def\eeqa{\end{eqnarray}}
\def\nn{\nonumber}
\def\barray{\left(\begin{array}}
\def\earray{\end{array}\right)}
\def\barraynb{\begin{array}}
\def\earraynb{\end{array}}

\def\?{\quad{\gb{\fbox{\texttt{?}}\;}}\quad}
\def\p{{\partial}}

\def\v0{\mathbf{0}}

\def\beq{\begin{equation}}
\def\eeq{\end{equation}}
\def\bea{\begin{eqnarray}}
\def\eea{\end{eqnarray}}

\def\p{\partial}

\def \p{{\partial}}

\def\6{\partial}
\def\7{\tilde}
\def\8{\widehat}

\def\G11{\Gamma_{11} }

\newcommand{\const}{\mathop{\rm const.}\nolimits}
\newcommand{\half }{\frac{1}{2}}

\def\smallover#1/#2{\hbox{$\textstyle\frac{#1}{#2}$}} %
\def\smallcirc{{\raise 0.5pt \hbox{$\scriptstyle\circ$}}}
\def\2{{\smallover1/2}}

\def\for{{\;\;\text{\small for}\;\;}}

\def\ie{{\;\text{\small i.e.}\;}}
\def\ie,{{\;\text{\small i.e.,}\;}}

\newcommand{\fm}{\mathfrak{m}}

\let\ssection=\section
\renewcommand{\section}{\setcounter{equation}{0}\ssection}

\begin{document}


\title{Displacement memory for flyby}

\author{
P.-M. Zhang$^{1}$\footnote{mailto:zhangpm5@mail.sysu.edu.cn},
Q.-L. Zhao$^{1}$\footnote{mailto: zhaoqliang@mail2.sysu.edu.cn},
J. Balog$^{2}$\footnote{mailto:balog.janos@wigner.hu},
P. A. Horvathy$^{3}$\footnote{mailto:horvathy@univ-tours.fr}
}

\affiliation{
${}^1$ School of Physics and Astronomy, Sun Yat-sen University, Zhuhai, China
\\
${}^{2}$ Holographic QFT Group, Institute for Particle and Nuclear Physics,
HUN-REN Wigner Research Centre for Physics
H-1525 Budapest 114, P.O.B. 49, Hungary
\\
${}^{3}$ Institut Denis-Poisson CNRS/UMR 7013 - Universit\'e de Tours - Universit\'e d'Orl\'eans Parc de Grammont, 37200; Tours, FRANCE 
\\
}
\date{\today}


\begin{abstract} 
Zel'dovich and Polnarev, in their seminal paper [1] extending previous work on the Memory Effect, suggested that a gravitational wave generated by flyby would merely displace the particles. We confirm their prediction numerically by fine-tuning the wave profile proposed by Gibbons and Hawking [2], and then analytically for its approximation by a P\"oschl-Teller potential. Higher-order derivative profiles proposed for gravitational collapse, etc [2] are shortly discussed.
\\

Annals Phys. \textbf{473} (2025), 169890.
doi:10.1016/j.aop.2024.169890
[arXiv:2407.10787 [gr-qc]]

\end{abstract}

\maketitle

\tableofcontents

\section{Introduction}\label{Intro}

The {Displacement Memory Effect} (DM) for \GWs  generated by flyby was proposed by Zel'dovich and Polnarev   
 \cite{ZelPol} (see also \cite{GibbHaw71,BraTho,BraGri,Sachs62,Christo}), extending previous work on the {Velocity Memory Effect} (VM) \cite{Ehlers,Sou73,GriPol}. 
Our previous investigations  
 \cite{ShortMemory,LongMemory,EZHRev} hinted, though,  at that for \emph{randomly chosen} parameters particles initially at rest  would fly apart with non-zero velocity, as illustrated, e.g. in FIG. \# 12. of ref. \cite{EZHRev}.
  However closer recent scrutiny \cite{DM-1, Jibril,ZZHflyby} indicates  that 
 for certain ``magical'' values of the wave parameters reminiscent of quantum numbers  we  \emph{do get DM}, highlighted by the vanishing of the velocity outside the Wavezone,
\beq
\frac{dX}{dU}\Big|(U=-\infty)=0=\frac{dX}{dU}\Big|(U=\infty)\,.
\label{DMcond}
\eeq 

This paper extends these  results to a more physical realm with particular attention at the analytically solvable 
\PT-version \cite{DM-1,ZZHflyby,PTeller,Chakra}. 
Then we generalize the flyby result to gravitational collapse, etc by studying  models with higher-order derivatives of the (Gaussian or \PT profile) proposed in \cite{BraTho,BraGri,GibbHaw71}. 
\goodbreak 

\smallskip
We start with the $4$ dimensional Brinkmann metric  \cite{Brink,DBKP,DGH91},
\beq
g_{\mu\nu}dX^\mu dX^\nu=
\delta_{ij} dX^i dX^j + 2 dU dV + 
\half{\cA}(U)\Big((X^1)^2-(X^2)^2\Big)dU^2\,,
\label{Bmetric}
\eeq
where $\bX=(X^i)$ are transverse coordinates and $U,\, V$ are light-cone coordinates. Its
geodesics are given by,
\begin{subequations}
\begin{align}
&\dfrac {d^2\!X^1}{dU^2}-\frac{1}{2}\cA X^1 = 0\,,
\label{geoX1}
\\[6pt]
&\dfrac {d^2\! X^2}{dU^2} + \frac{1}{2}\cA X^2 = 0\,,
\label{geoX2}
\\[8pt]
&
\dfrac {d^2\!V}{dU^2} +\frac{1}{4}\dfrac{d\cA}{dU}\Big((X^1)^2-(X^2)^2\Big) 
+ 
\cA\Big(X^1\frac{dX^1}{dU}-X^2\dfrac{dX^2}{dU}\Big)=0\,.
\label{geoV}
\end{align}
\label{Bgeoeqn}
\end{subequations}
\noindent 
The last equation is solved  by solving first eqns \eqref{geoX1} - \eqref{geoX2} for the transverse trajectory $\bX(U)$ and then lifting it to the \GW spacetime.

Let us recall that geodesic motion posses a conserved quantity referred to as the Jacobi invariant \cite{Eisenhart,EDAHKZ,SchRev},
\beq
{\fm}^2 = -g_{\mu\nu}\dot{X}^{\mu}\dot{X}^{\nu} = \const 
\label{Jacobiinv}
\eeq
where the dot denotes $d/dU$.
Discarding tachions, we shall consider ${\fm}^2 \leq0$.

Postponing the study of the  $V$-motion to Sec.\ref{Vmotion}, we first focus our attention at the motion in transverse space. 
FIG.\ref{Gaussd1Geo} suggests that for a random choice of the parameters the outgoing particle has constant nonzero velocity~: it exhibits the \emph{velocity effect} (VM). However  recent results  \cite{DM-1} suggest that for a judicious choice of the amplitude a ``miracle" may happen and the outgoing velocity \emph{can} indeed vanish --- letting us wonder if this is also true in a broader and more physical context, and  for flyby  in particular. Our investigations below confirm that this is indeed what happens.

In our previous paper \cite{DM-1} we studied two cases in detail, namely those whose Brinkmann profile $\cA$ is either a simple Gaussian, \eqref{AGauss}, or a \PT potential, \eqref{APT}. Both of them have an integer wave number $m$.
In Sects.\ref{flybyDM} and \ref{dPTsec} we study what happens 
for flyby modelled by the \emph{derivative profiles} \cite{GibbHaw71} \eqref{A1G} and
\eqref{A1PT}, respectively.

\section{Flyby: transverse motion}\label{flybyDM}

\subsection{Flyby profile: derivative of Gaussian}\label{dGsec}

\begin{figure}[h]
\includegraphics[scale=.185]{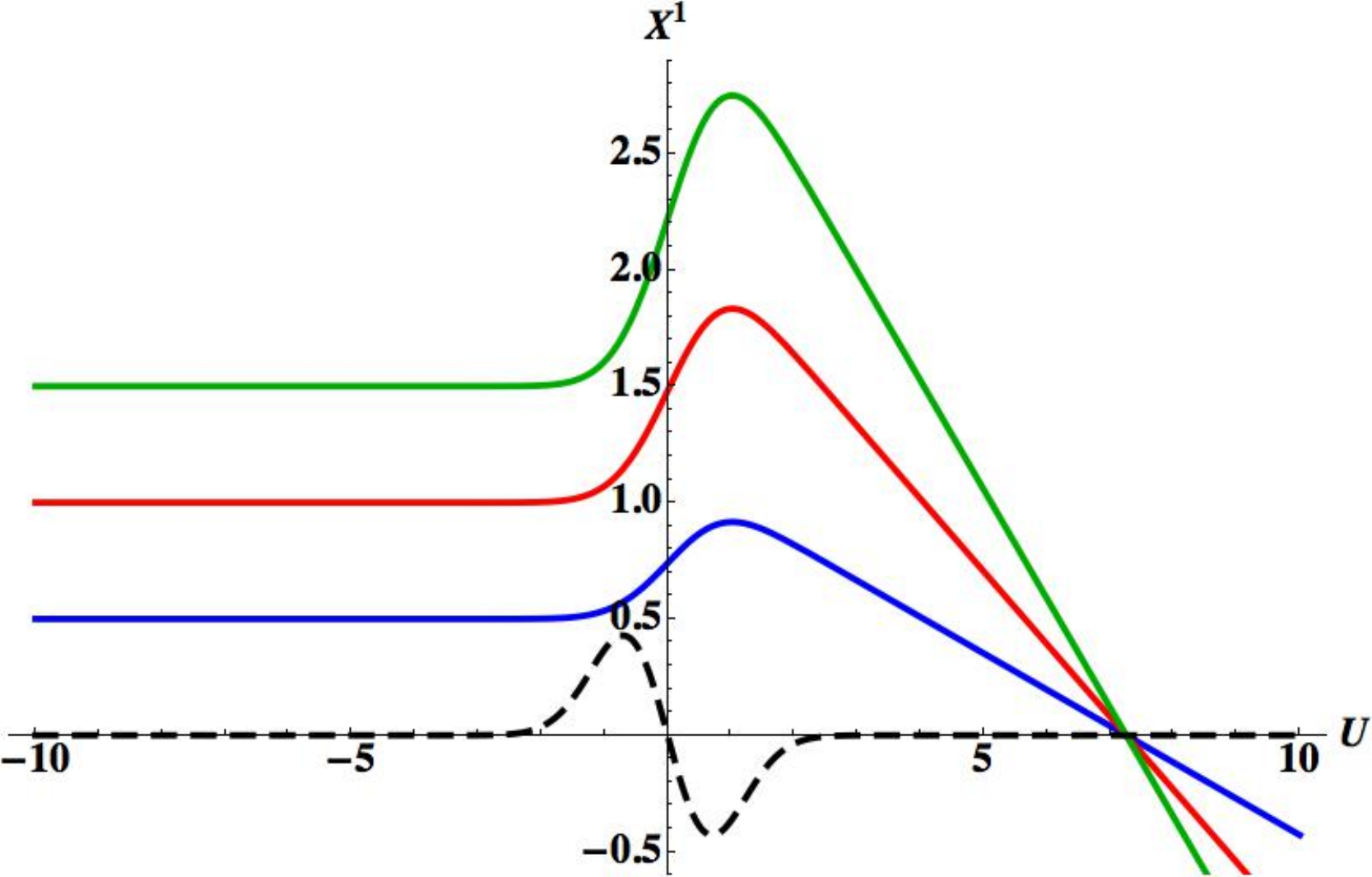}\,
\includegraphics[scale=.185]{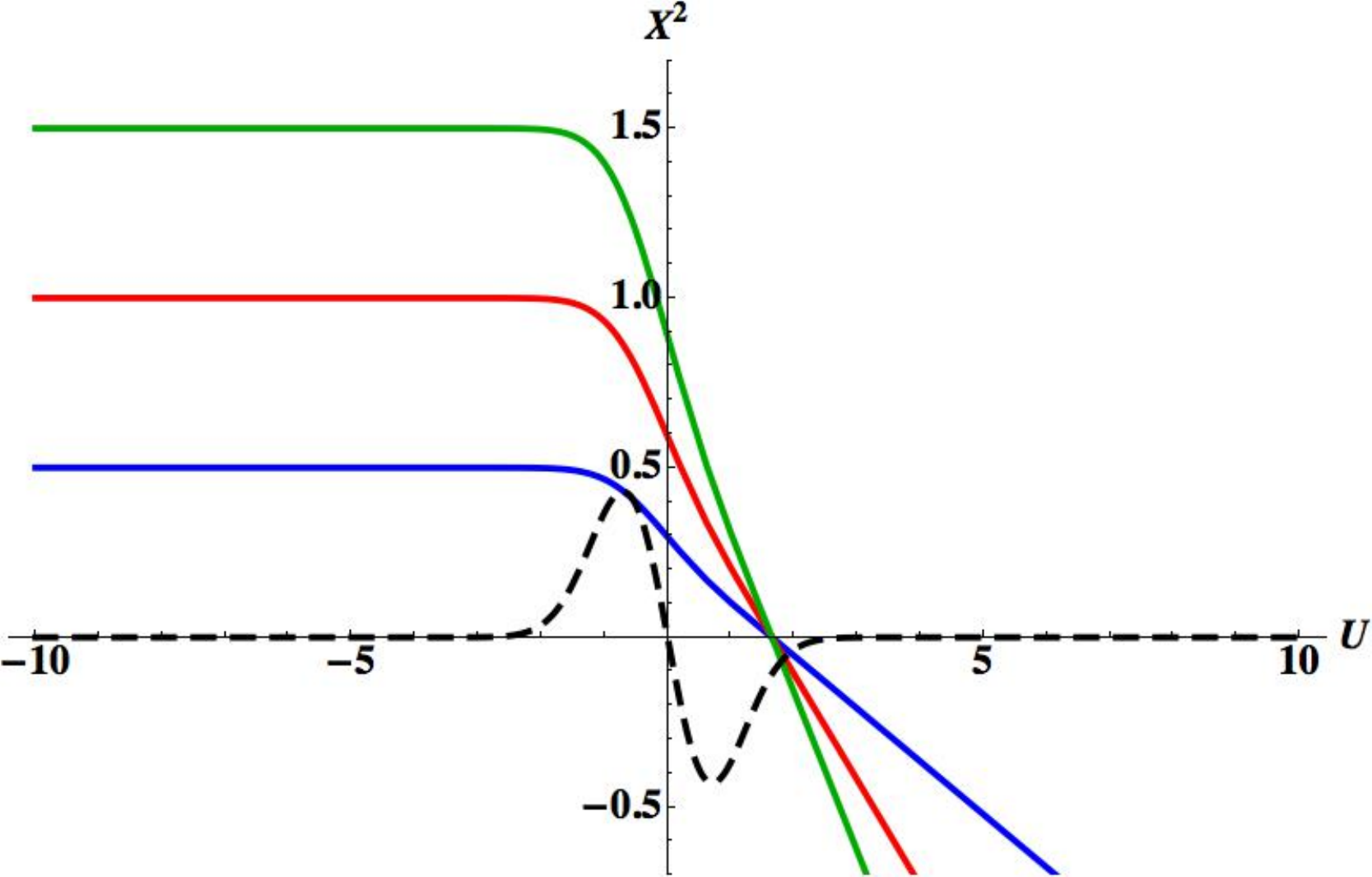}\,
\includegraphics[scale=.185]{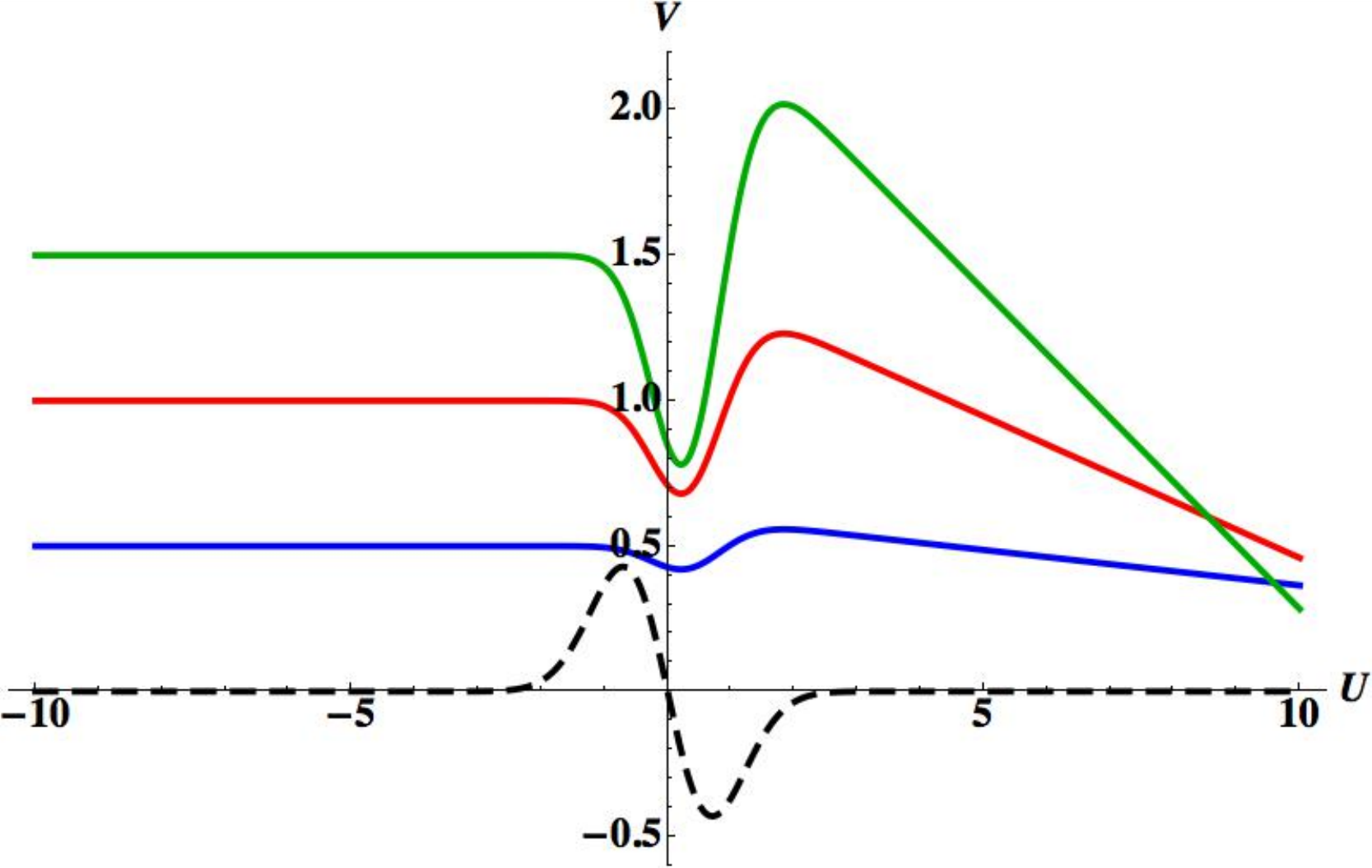}\\
\caption{\textit{The null geodesics for the first derivative of the Gaussian  in \eqref{A1G} shown in the background exhibit, for randomly choosen amplitude $k$, the velocity effect (VM).}
\label{Gaussd1Geo}
}
\end{figure}

The profile for flyby proposed by Gibbons and Hawking in \cite{GibbHaw71}  is,
\begin{equation}
\cA \equiv \cA^{G}=
\frac{\;\;\,d}{dU}\left(\frac{k}{\sqrt{\pi}}%
e^{-U^{2}}\right)\,.
\label{A1G}
\end{equation}%
 Its null geodesics are depicted in FIG.\ref{Gaussd1Geo},

\begin{figure}[h]\hskip-2mm
\includegraphics[scale=.36]{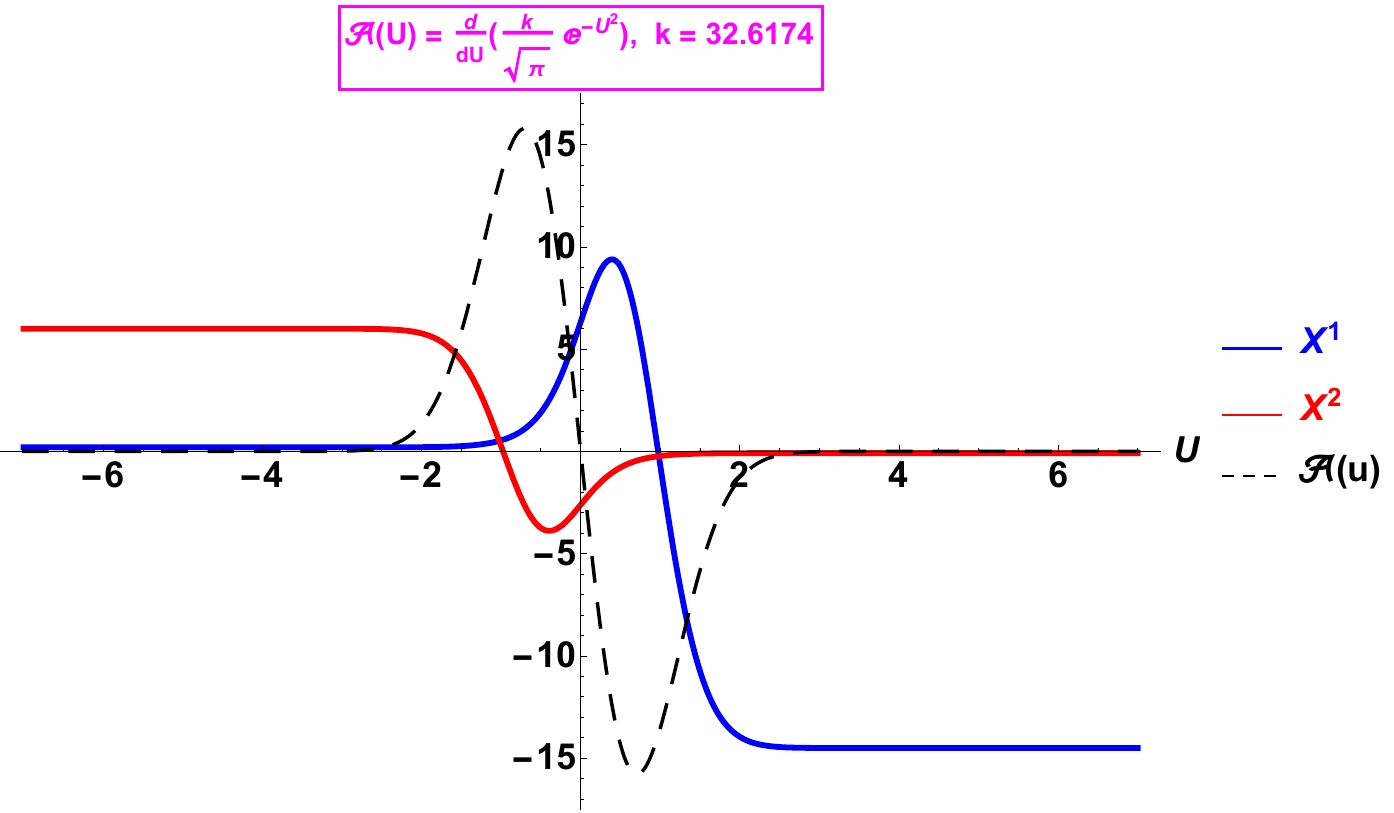}
\\[4pt]
\includegraphics[scale=.36]{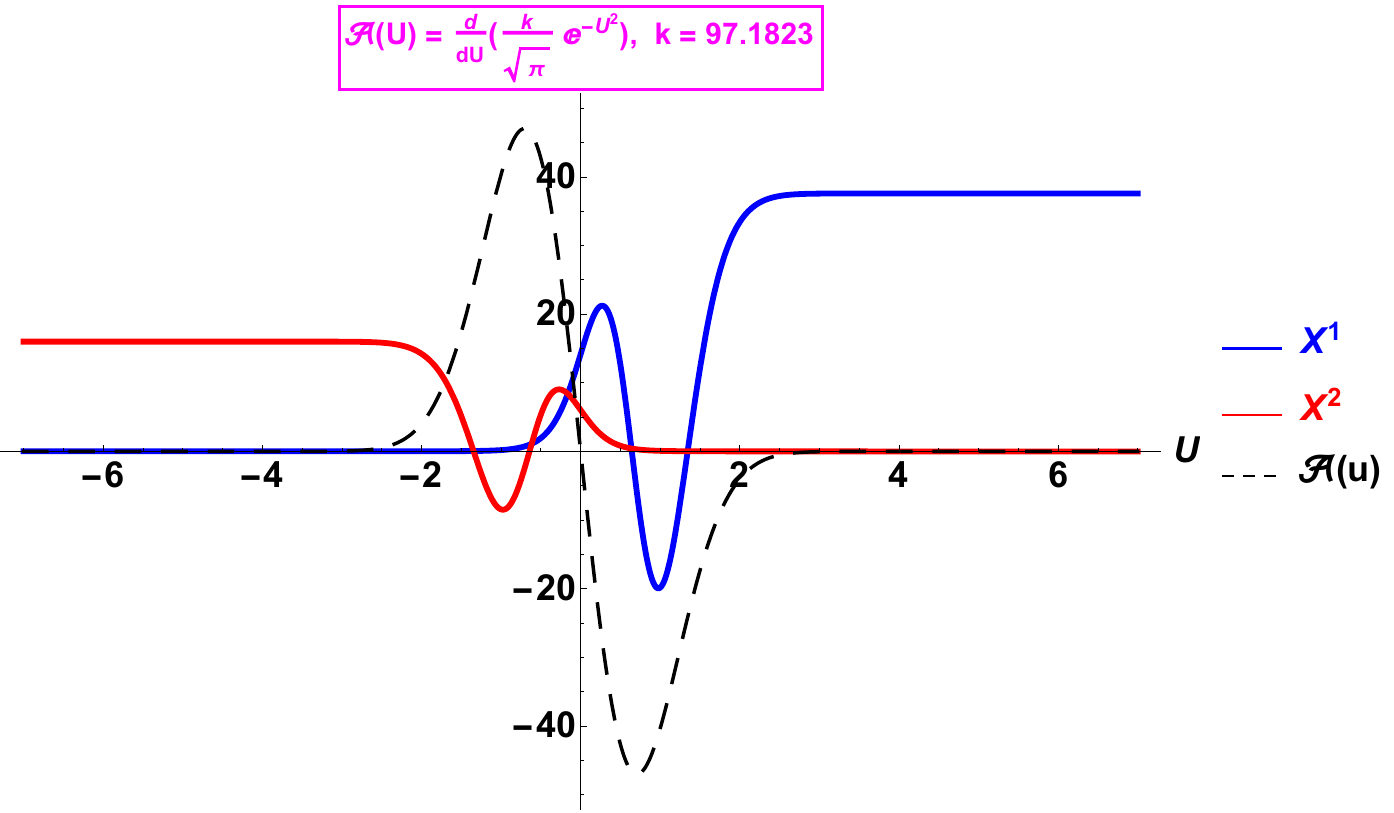}
\vskip-4mm
\caption{\textit{For the flyby profile \eqref{A1G} 
 fine-tuning the amplitude yields, for each $k=k_{crit}$, a unique (approximate) DM trajectory for \underline{both} components, as shown for 
 \magenta{${\bf m=1}$} and  \magenta{${\bf m=2}$}.
}
\label{d1-Gauss-m12}
}
\end{figure}

We start with the massless case, $\fm=0$. 
At the first sight, the null geodesics show VM as seen in FIG.\ref{Gaussd1Geo}. 
We study first the transverse motion; that in the $V$-direction will be considered in sect.\ref{Vmotion}.
Numerical fine-tuning then indicates, unexpectedly, that for certain ``magical'' critical values  $k=k_{crit}$ \emph{we do get (approximate) {DM}} for both transverse coordinates, as illustrated  in FIG.\ref{d1-Gauss-m12}. In fact
 we found a well separated discrete series of critical amplitudes, $k_m$, each of which corresponding to a unique DM wave labeled by an integer $m$, 
\beq
k_{1}=32.6, 
\quad
k_{2}=97.2,
\quad
k_{3}=194.8,
\quad
k_{4}=325.5,
\quad
k_{5}=489.2
\;
\dots
\label{Gcritks}
\eeq
Our plots suggest that for each such critical amplitude $k_{crit}$ the Wavezone contains a unique {DM} trajectory composed approximately of an integer number of half-waves, consistently with our observations  in \cite{DM-1}.
The novelty is that DM is now obtained for \emph{both} coordinates, whereas for the simple Gaussian, shown in FIG.20 of \cite{DM-1}, DM could be found in the attractive, but not in  the repulsive sector; in \cite{DM-1} we called this ``half-DM''.

\begin{figure}[h]
\includegraphics[scale=.33]{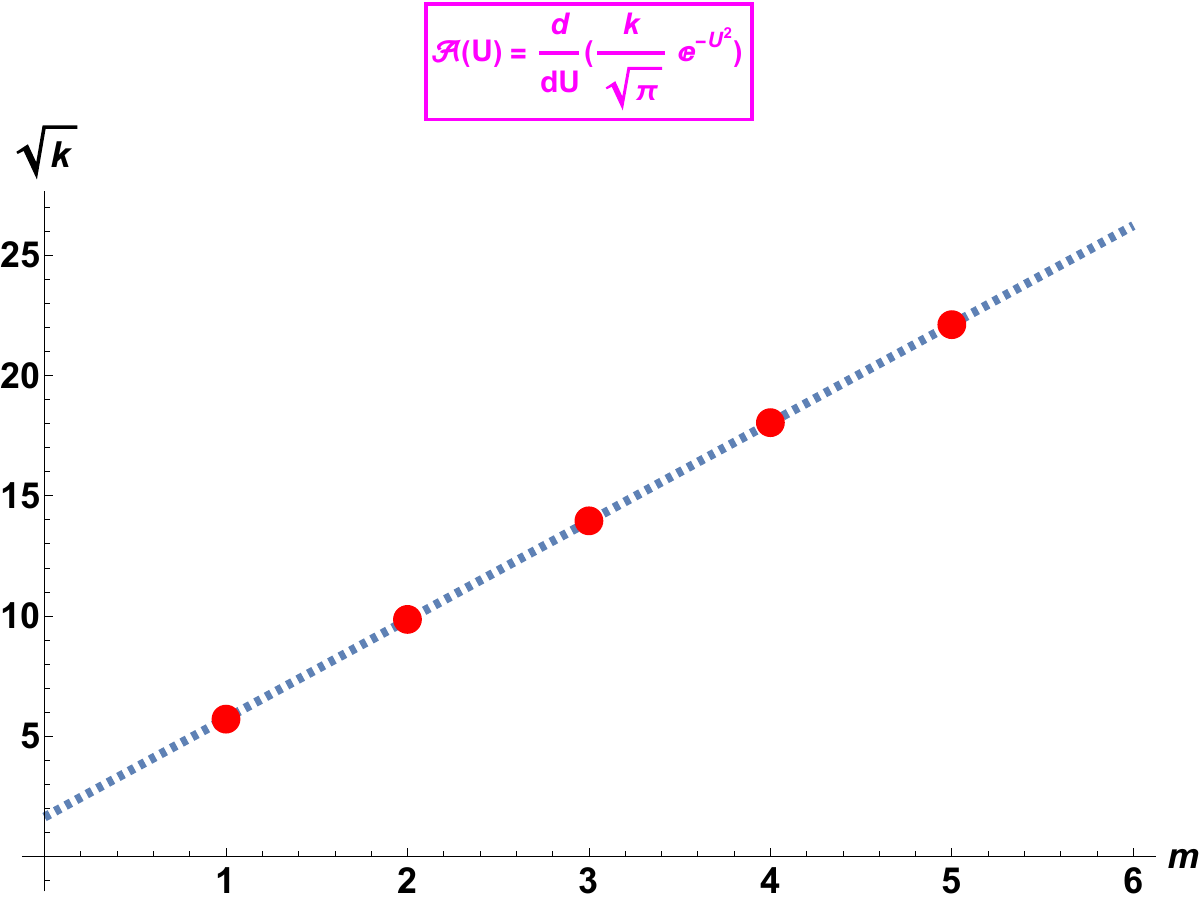}
 \vskip-3mm\caption{\textit{\small  DM is obtained when the wave number \magenta{$\bf m$} and the square-root of the amplitude 
\orange{$\bf \sqrt{k}$} are (approximately) linearly correlated. 
}
\label{fly-mk1}
}
\end{figure}

The [square-roots of] the critical amplitude, $\sqrt{k_{crit}}$, depend on $m$  approximately linearly, as shown in FIG.\ref{fly-mk1} (to be compared to FIG.5 of \cite{DM-1}). 

For $k\neq k_{crit}$ we get instead non-zero outgoing velocity; the trajectory is not composed of an integer number of half-waves: we get VM but no DM, as it is manifest in FIG.\ref{Gaussd1Geo}.

\begin{figure}[h]
\includegraphics[scale=.3]{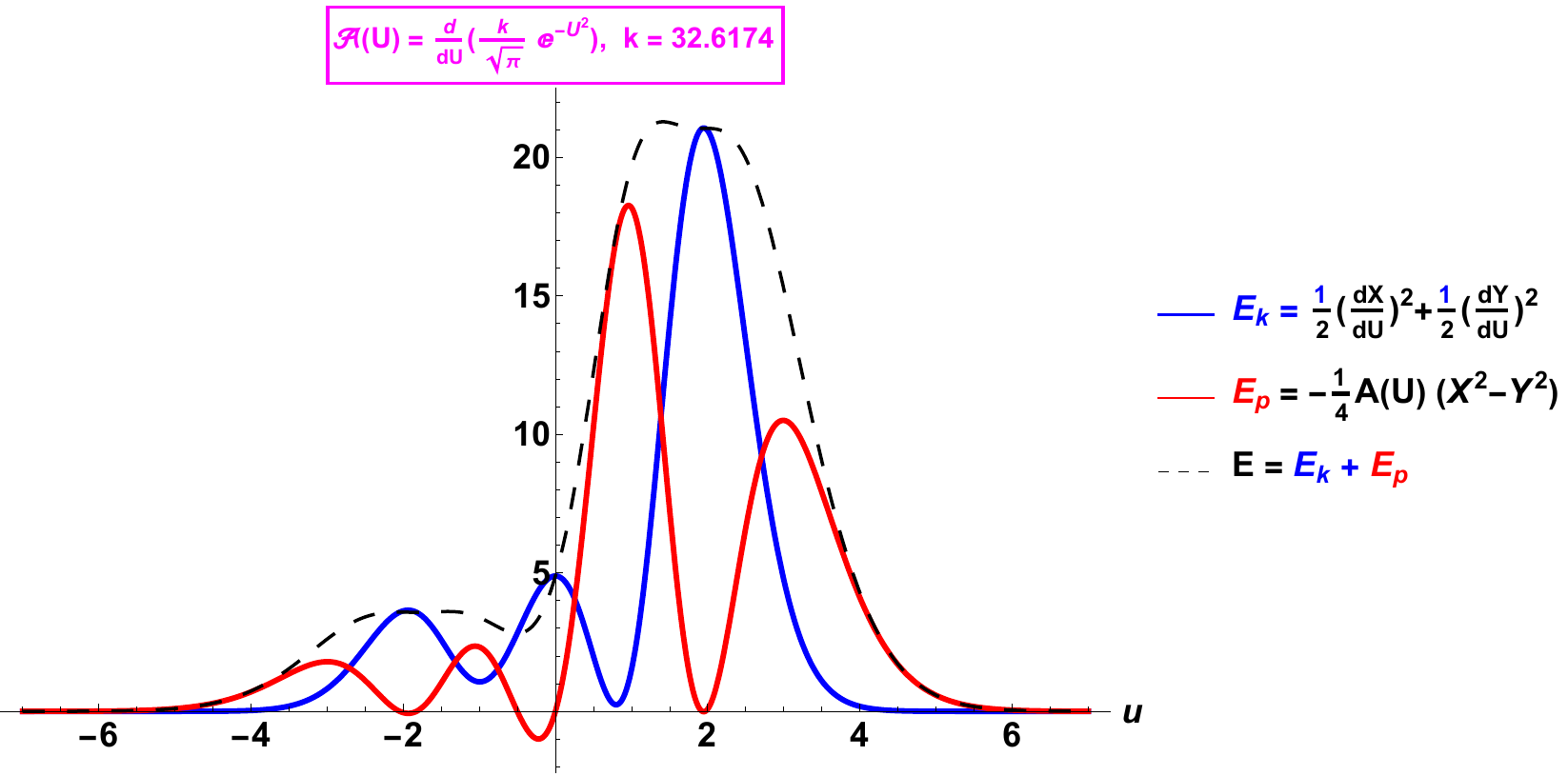}
\hskip-19mm
\includegraphics[scale=.335]{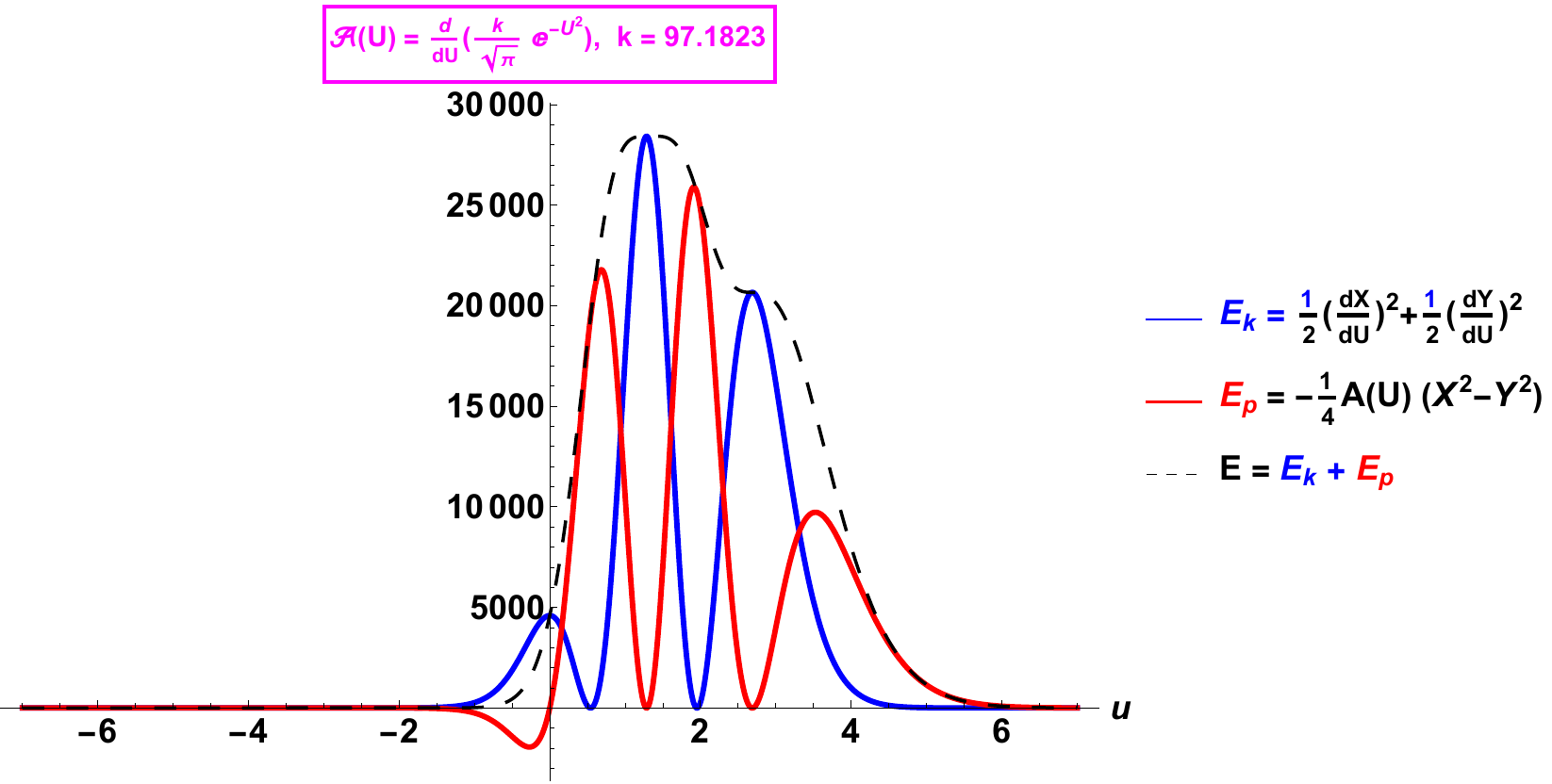}
\vskip-4mm
\caption{\textit{For $k=k_{crit}$ the total energy balance after the flyby is zero, as shown for \magenta{${\bf m=1}$} and for \magenta{${\bf m=2}$}.
} 
\label{flybyenergym12}
}
\end{figure}

For generic wave parameters, the particle absorbs some energy from the passing wave \cite{Bondi57,Ehlers,exactsol}.
DM implies in turn zero cumulated energy   
of the solution  in FIG.\ref{d1-Gauss-m12}, as confirmed in FIG.\ref{flybyenergym12}.
\goodbreak
\goodbreak

\begin{figure}[h]
\includegraphics[scale=.395]{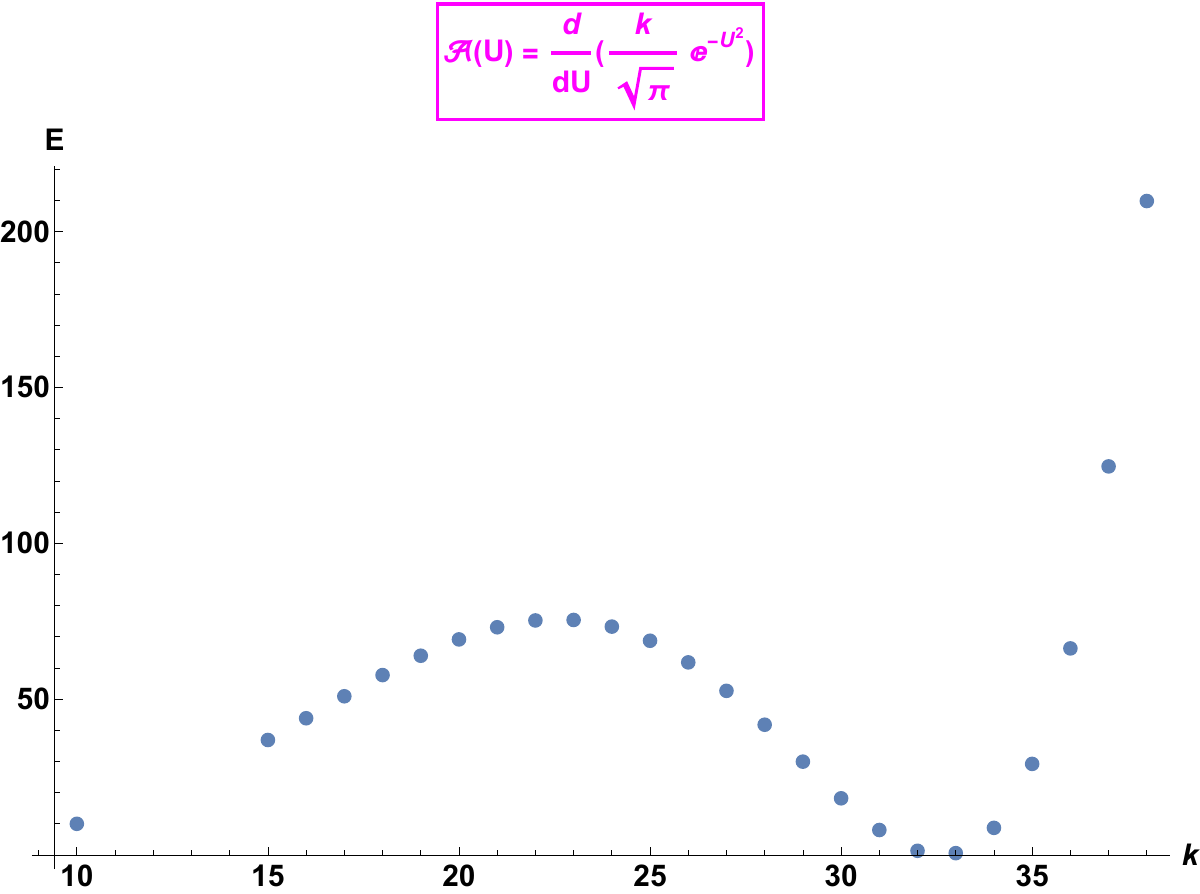}\,
\includegraphics[scale=.385]{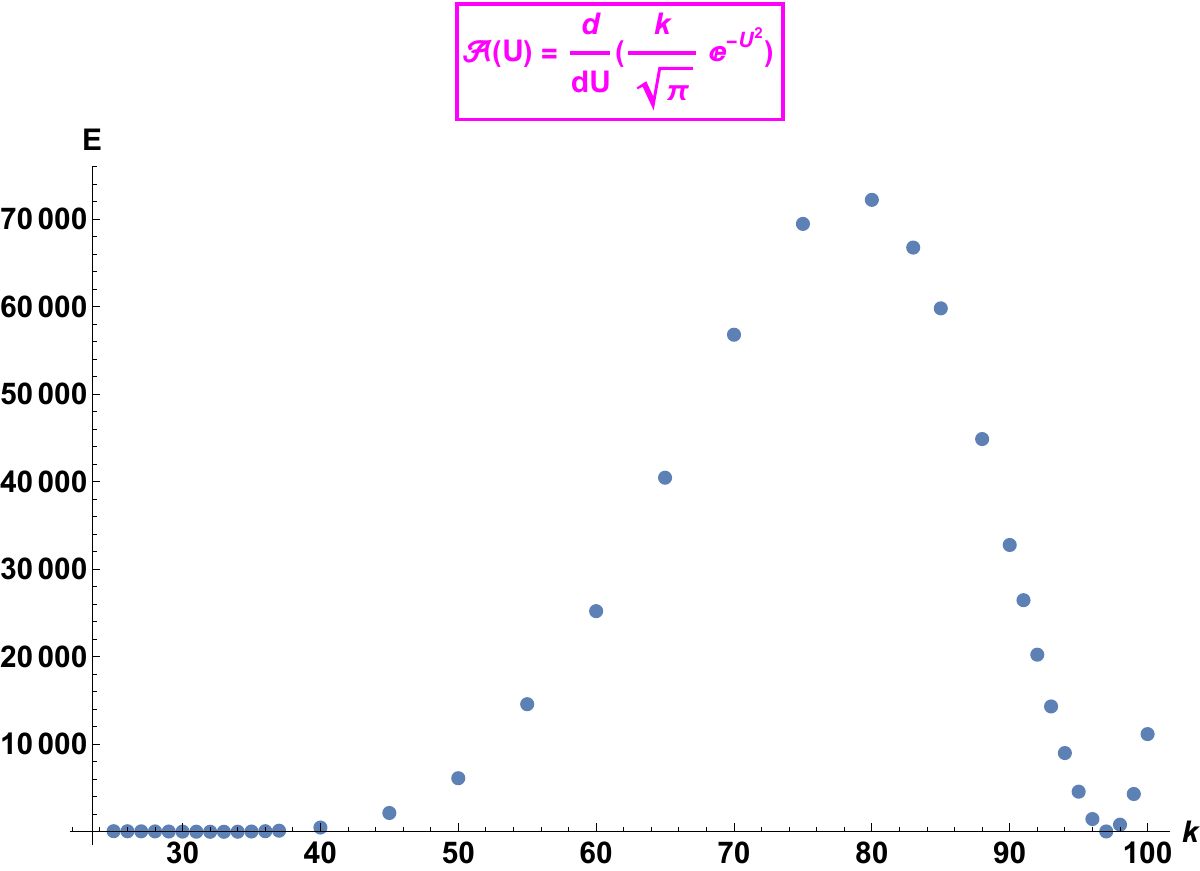}\vskip-3mm	
\caption{\textit{For  $k\neq k_{crit}$
 we have VM and the particle extracts some energy from the passing \GW\!. In the fine-tuned case $k=k_{crit}$, though, the cumulative energy balance is zero: the particle does not absorb any energy.
 }
\label{smalllargek}
}
\end{figure}

 It is instructive to record the energy absorption of the underlying NR model \cite{Eisenhart,DBKP,DGH91} for $k_{crit}$ both in the low and high amplitude regimes, shown in FIG.\ref{smalllargek}.
We shall return to the relation to the underlying NR system in sec.\ref{Vmotion}.

\section{\PT model for flyby: numerical study}\label{dPTsec}

First we revisit the simplest cases.
The geodesics for the  Gaussian profile proposed in \cite{GibbHaw71} 
\begin{equation}
\cA^G(U) = \frac{k}{\sqrt{\pi}}e^{-U^{2}}
\label{AGauss}
\eeq 
can only be found numerically. However
its similarity with the P\"oschl-Teller profile  \cite{PTeller},
\begin{equation}
\cA^{PT}(U) = \dfrac{k}{2\cosh^2 U}\,,
\label{APT}
\end{equation}%
 allowed us to find analytic solutions by solving the \PT\!-modified 
 Sturm-Liouville equation \cite{DM-1,Jibril,Chakra,ZZHflyby} 
\beq
\dfrac {d^2\! X}{dU^2} + \frac{1}{2}\cA^{PT} X = 0\,.
\label{PTX2}
\eeq
Putting $t=\tanh U$ and introducing the 
(at this point real) parameter $m$  by
\beq
k= 4\, m(m+1) 
\label{PTkm}
\eeq 
\eqref{PTX2} then becomes the Legendre equation,
\beq
(1-t^2)\frac{d^2X}{dt^2}-2t\frac{dX}{dt} + m(m+1)\, X = 0\,.
\label{PTSL}
\eeq

Our clue is that the DM conditions \eqref{DMcond} [or $X(U=\mp\infty)\to\const$] require that the solution extends to $t= \mp1$, --- yielding a \emph{quantization condition}: the parameter $m$ must be a natural number, 
\beq
m = 1, \, 2,\, \dots 
\eeq
Then the Legendre \emph{function} which solves eqn. \eqref{PTSL} truncates to a Legendre \emph{polynomial} of order $m$, $P_m(\tanh U)$, see eqn. \#(IV.6)  and the trajetories exhibit DM, as illustrated in FIGs. \# 7, 8, 9 of \cite{DM-1}.
In a remarkable analogy with what happens for a harmonic oscillator in Quantum Mechanics \cite{DM-1},  $m\geq 1$ counts  \emph{the number of half-waves accommodated in the (approximate) wavezone}.
The question will be further discussed in sect.\ref{analSec}. 
 
In $D=2$ transverse dimensions the two components come with a relative minus, therefore the two equations in \eqref{geoX1}-\eqref{geoX2} differ by the sign of $k$.
Bounded solutions arise only in the attractive sector -- Legendre polynomials have $m \geq 1$ -- and we get  ``half-DM'' (as in the Gaussian case \cite{DM-1}). 

\medskip
Turning to \emph{flyby} which is our main concern in this paper, we note that the derivative of the \PT potential (dPT), 
\begin{equation}
\cA \equiv \cA^{dPT}=
\frac{\;\;\,d}{dU}\left(
\frac{k}{2\cosh^2U}\right)
=  -k \Big(\tanh U-\tanh^3 U\Big)\,
\label{A1PT}
\eeq
shown in FIG.\ref{dGdPT} is reminiscent of the derivative-Gaussian proposed in \cite{GibbHaw71}, prompting us to study the geodesics of the metric \eqref{Bmetric} with 
 \emph{derivative \PT} (dPT) profile \eqref{A1PT} as an alternative model for flyby. 
\begin{figure}[h]
\includegraphics[scale=.7]{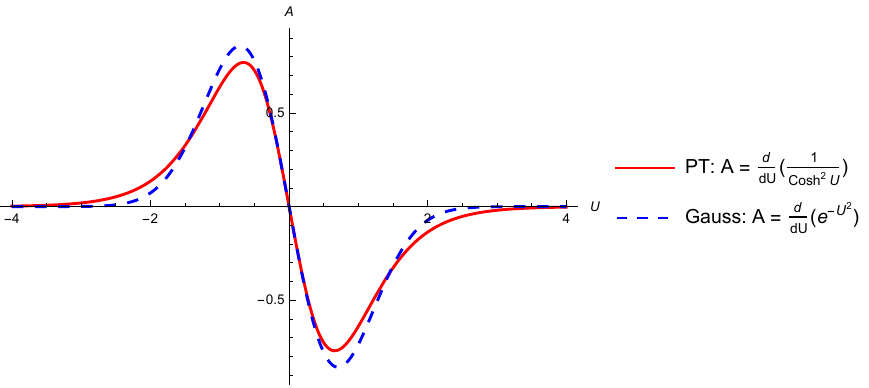}\vskip-4mm\caption{\textit{\small The derivative of the \PT potential is a good approximation of that of the Gaussian.
}
\label{dGdPT}
}
\end{figure}

Then we get, instead of  \eqref{PTSL},  %
\beq
(1-t^2)\frac{d^2X^i}{dt^2}-2t\frac{dX^i}{dt\;} \pm
\frac{k}{2} \, t \,X^i 
=0\,,
\label{dPTeq}
\eeq 
with the $\pm$ sign referring to the coordinate sector,  $i=1$ or $2$. Assuming $k>0$, 
 $+k$ is for $X^2$ and  $-k$ is for $X^1$.
 
The behavior is reminiscent of  but  different from that for simple \PT\!,  \eqref{PTSL}, as seen by decomposing \eqref{dPTeq} as,
\beq
\underbrace{(1-t^2)\frac{d^2X^i}{dt^2}-2t\frac{dX^i}{dt} 
+ 2m(m+1) \,X^i}_{Legendre} \quad - \quad 
\underbrace{
 (1\mp t)
 \,2m(m+1) \,X^i}_{perturbation}
=0\,.
\label{LegPerturb}
\eeq
where $m$ is an {\rm a priori} \emph{real} number. Thus \eqref{dPTeq} is  a Legendre eqn. with a linear perturbation.
For $t \to \pm1$  both equations become Legendre equations but with different coefficients of the linear term, see
FIGs. \ref{d1-PT-m12}~and~\ref{dPT166-276}.

We emphasise, though, that while for the PT profile the analytic calculation is routine-like, the corresponding one for dPT is highly involved, as it will be seen in detail in sect.\ref{analSec}.
\goodbreak

\begin{figure}[h]\hskip-3mm
\includegraphics[scale=.33]{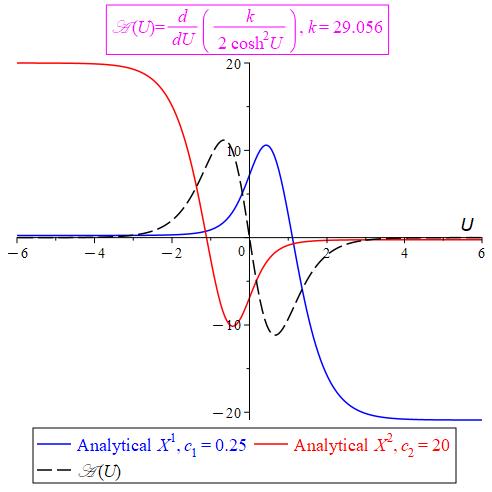}\qquad
\includegraphics[scale=.33]{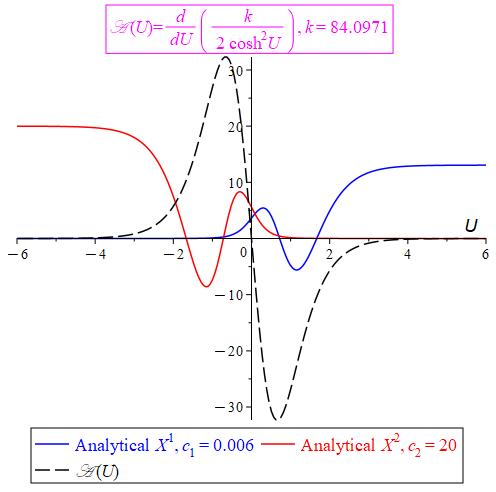}
\\{}\vskip-3mm\hskip-3mm
\magenta{${\bf m=1}$}
\hskip56mm
\magenta{${\bf m=2}$}\\
\vskip-4mm
\caption{\textit{For the dPT-flyby profile \eqref{A1PT}, both numerically or analytically obtained trajectories exhibit DM for {both} components.
For convenience, the $X^2$ solutions (in \red{red}) are rescaled by a factor 20, while the $X^1$ solutions (in \blue{blue}) are rescaled, for \magenta{${\bf m=1}$}, \magenta{${\bf m=2}$}, by 0.25 and 0.006, respectively.   
}
\label{d1-PT-m12}
}
\end{figure} 
\goodbreak
%
\begin{figure}[h]
\includegraphics[scale=.33]{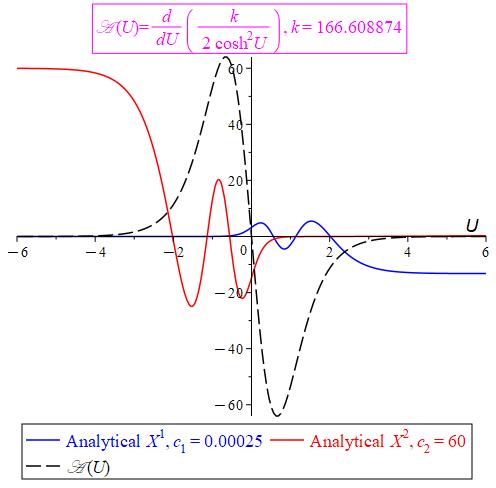}\qquad
\includegraphics[scale=.33]{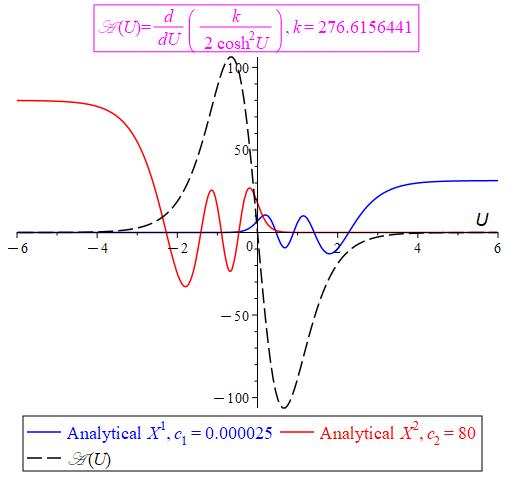}
\\{}\vskip-3mm
\hskip1mm
\magenta{${\bf m=3}$}\hskip55mm
\magenta{${\bf m=4}$}\\ \vskip-4mm
 \caption{\textit{\small 
 Analytic dPT trajectories for  critical amplitude (i) $k_{crit}=166.\dots$  and (ii) $k_{crit}=276.\dots$  in the derivative-\PT approximation, have half-wave numbers \magenta{${\bf m=3}$} and  \magenta{${\bf m=4}$}. }
\label{dPT166-276}
}
\end{figure}
\goodbreak

We note for completeness that off the critical values, $k\neq k_{m}$, the trajectories exhibit the velocity effect (VM) as expected 
and illustrated in FIG.\ref{A-dPT-30}. 

\begin{figure}[h] 
\includegraphics[scale=.27]{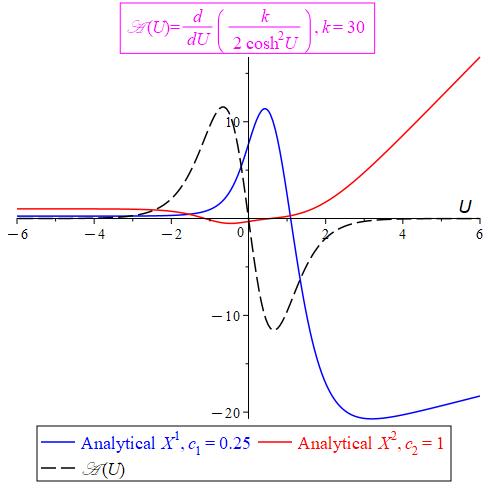}\qquad\qquad
\includegraphics[scale=.26]{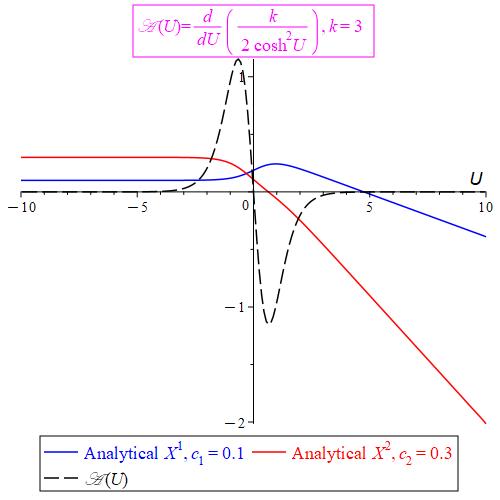}
\vskip-3mm
\caption{\textit{\small For \orange{${\bf k\neq k_{crit}}$} 
 the trajectories exhibit the velocity effect (VM).
 }
\label{A-dPT-30}
}
\end{figure}
\goodbreak

\section{\PT model for flyby: (semi-) analytic treatment}\label{analSec}

Using  $t=\tanh U$  the equations for the transversal components are \eqref{dPTeq},
\begin{equation}
(1-t^2)\frac{{\rm d}^2 X^i}{{\rm d}t^2}-2t\frac{{\rm d}X^i}{{\rm d}t}
\pm\frac{k}{2}tX^i=0\,, 
\nn     
\end{equation}
 where the $\pm$ sign is for the $i=1,2$ components, respectively. We introduce the parameter
\begin{equation}
p=\mp\frac{k}{2}
\end{equation}
and allows us to study the two cases simultaneously. Changing the independent
variable to
$ 
z=\frac{t+1}{2}
$ 
the equation to be solved becomes,
\begin{equation}
\frac{{\rm d}^2X}{{\rm d}z^2}
+\left(\frac{1}{z}+\frac{1}{z-1}\right)\frac{{\rm d}X}{{\rm d}z}
+p\left(\frac{1}{z}+\frac{1}{z-1}\right)X=0,
\end{equation}
where $X=X^1$ or $X^2$. Here we recognize the \emph{confluent Heun equation} with special
parameters. The singular points are $z=0,1$ (plus the irregular singular
point at infinity), corresponding to $t=-1,1$ and $U=-\infty,+\infty$,
respectively. 

We are looking for solutions which are regular at both of these points.
Assuming regularity at $z=0$, we can write the solution as a power series
\begin{equation}
X=\sum_{n=0}^\infty s_nz^n.
\label{power}
\end{equation}
We normalize the solution by putting
$ 
s_0=1.
$ 
The $s_n$ coefficients can be determined from the recursion relation
\begin{equation}
s_{n+1}=\frac{[n(n+1)-p]s_n+2ps_{n-1}}{(n+1)^2},\qquad n=0,1,\dots
\end{equation}
($s_{-1}=0$ by convention.)

We look for parameter values such that the power series (\ref{power}), which 
converges in the disk $|z|<1$, defines a function which is regular also at $z=1$. In
similar problems the solution is usually a polynomial.
There is no polynomial solution
in this case, though.

In the original variable the general solution of the problem  behaves as
\begin{equation}
X\sim 2A_0 U+{\rm const.}
\label{Xinf}
\end{equation}
for $U\to+\infty$, where $A_0(p)$ is some parameter-dependent coefficient.  
In the new coordinates this means singular behaviour,
\begin{equation}
X\sim -A_0\ln\left(\frac{1-t}{2}\right)=-A_0\ln(1-z)
=A_0\sum_{n=1}^\infty\frac{z^n}{n}
\end{equation}
for $t\to1$, $z\to1$.
We introduce
\begin{equation}
x_n=ns_n
\end{equation}
for convenience. According to the above analysis, this has asymptotic
expansion for large $n$
\begin{equation}
x_n=A_0+\frac{A_1}{n}+\frac{A_2}{n(n-1)}+\frac{A_3}{n(n-1)(n-2)}+\dots
\label{asy}
\end{equation}
with $p$-dependent coefficients $A_k$. Actually it follows from the recursion relation
that all higher coefficients are proportional to $A_0$:
\begin{equation}
A_k=\frac{a_k}{k!}A_0,
\end{equation}
where
\begin{equation}
a_1=-p,\quad a_2=p(p-2),\quad p_3=-p^2(p-8),\quad p_4=p^3(p-20),\quad{\rm etc.}
\label{Ak}
\end{equation}
We look for non-singular solutions with $A_0$=0 for
which $X(U)$ goes to a mere constant when $U\to\infty$, as it follows from \eqref{Xinf}. According to the above
considerations in this case all $A_k$ vanish and such solutions are not only
regular at $z=1$ but also analytic there. In fact, these solutions are entire
functions.

\subsection{Results}

It remains to find special parameter values $p_m$ such that 
\begin{equation}
A_0(p_m)=0\,.
\label{A00}
\end{equation}
(We will also use $k_m=2p_m$.)
\begin{figure}[h]
	\centering
	\includegraphics[width=.33\textwidth]{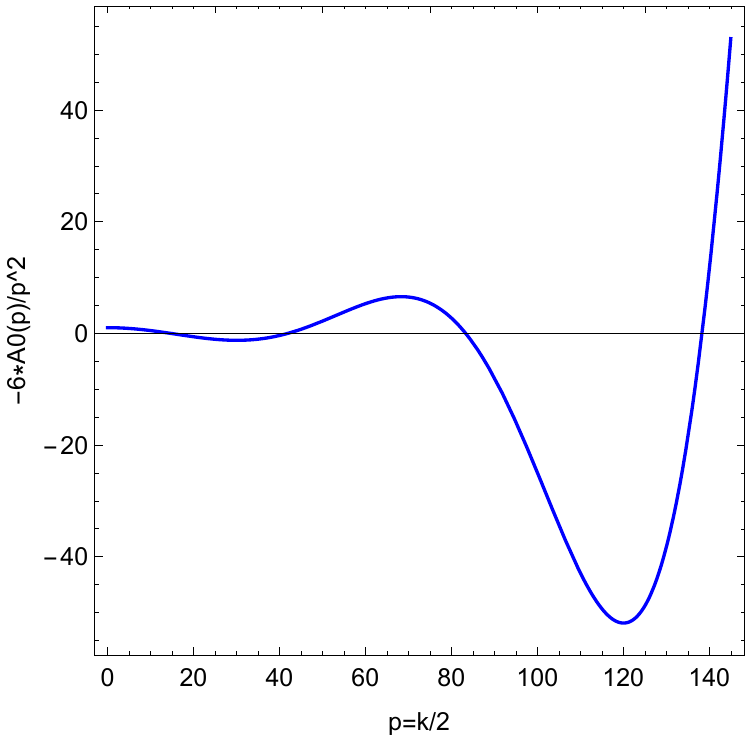}
	\caption{The rescaled $A_0(p)$ coefficient}
	\label{fig:FO4}
\end{figure}
The trick, which is often used in the case of series behaving asymptotically
like (\ref{asy}), is to apply a series of Robertson transformations. These
are defined recursively as
\begin{equation}
x^{(0)}_n=x_n,
\end{equation}
\begin{equation}
x^{(s)}_n=\frac{1}{s}\left\{(n+1-s)x^{(s-1)}_n-(n+1-2s)x^{(s-1)}_{n-1}\right\},
\qquad s=1,2,\dots
\end{equation}
This series of transformations eliminates, step by step,
the leading corrections such that for large $n$
\begin{equation}
x^{(s)}_n=A_0+{\rm O}\left(\frac{1}{n^{s+1}}\right),
\end{equation}
i.e. the $s^{\rm th}$ Robertson transformed series has the same leading term
but much smaller corrections to it.
Using this method we calculated (see FIG.\ref{fig:FO4})
\begin{equation}
A_0(p)= -\frac{p^2}{6}\left\{1-\frac{p^2}{180}+4.036062\cdot 10^{-6}p^4-
7.801993\cdot 10^{-10}p^6+\dots\right\}
\end{equation}

The function is even in $p$ which means that the same critical values work for
both signs in the original equation, i.e. the DM condition is satisfied
simultaneously for both components $X^1$, $X^2$.
The roots of $A_0(p)$ can only be calculated numerically \footnote{The critical values could also be derived readily by requiring that  the velocity  go to zero when $U \to \infty$, \eqref{DMcond}.
}
\begin{equation}
\begin{split}
k_1=29&.0560068187,\\
k_2=84&.0970923581,\\
k_3=166&.6088726851,
\end{split}
\qquad
\begin{split}
k_4&=276.6156438862,\\
k_5&=414.1212560177,\\
k_6&=579.1267981111.
\end{split}
\end{equation}

\subsection{Asymptotic values}

We normalize our solution such that $X=1$ at $U=-\infty$ ($z=0$). For critical
values of $k$ we can calculate $X$ at $U=+\infty$ ($z=1$). Let us take first $k=k_1$.
The solution corresponds to the $X^2$ component and we find
\begin{equation}
X^2(\infty)=-0.0119701374443\,.
\end{equation}
which is very nearly zero.
For $k=-k_1$ instead, which corresponds to the $X^1$ component, we find the considerably larger value,
\begin{equation}
X^1(\infty)=-83.5412300512\,.
\end{equation}
Of course, $X^1(\infty)X^2(\infty)=1$, due to the $U\leftrightarrow -U$
symmetry of the problem and our normalization.
For $k=\pm k_2$ the analogous results are
\begin{equation}
X^1(\infty)=2176.4880022386,\qquad
X^2(\infty)=0.000459455884207\,.
\end{equation}

\subsection{The $k_m\leftrightarrow m$ relation}

The following, purely numerically obtained  formula,
\begin{equation}
k_m=b_0 w_m+b_1+b_2/w_m\,,
\end{equation}
where $w_m=m(m+1)$ 
and
\begin{equation}
b_0=13.750365230964,\qquad
b_1=1.614330366353,\qquad
b_2=- 0.117986675402.   
\end{equation}
works with rather good precision.

%

\subsection{Coefficient of the $\ln(1-z)$ singularity}

Summation of the $A_0$ term gives
\begin{equation}
\sum_{n=1}^\infty \frac{z^n}{n}=-\ln(1-z)\,.
\end{equation}
The higher terms proportional to $A_1$, $A_2$, etc. all contain terms
proportional to $\ln(1-z)$, but these terms are not singular in the sense
that they have finite $z\to1$ limits. Let us parametrize these higher terms as
\begin{equation}
\sum_{n=k}^\infty \frac{z^n}{n^2(n-1)(n-2)\dots(n-k+1)}=
-c_k\ln(1-z)+r_k.
\end{equation}
For example,
\begin{equation}
c_1=-\zeta-\frac{\zeta^2}{2}-\frac{\zeta^3}{3}+\dots\qquad
r_1=\quad \frac{\pi^2}{6}-\zeta-\frac{\zeta^2}{4}-\frac{\zeta^3}{9}+\dots \quad
\end{equation}
\begin{equation}
c_2=\frac{\zeta^2}{2}+\frac{\zeta^3}{3}+\dots\qquad\qquad\; r_2=2-\frac{\pi^2}{6}-\zeta+\frac{\zeta^2}{4}+\frac{\zeta^3}{9}+\dots
\end{equation}
Here for convenience we introduced the new variable
\begin{equation}
\zeta=1-z.
\end{equation}
The coefficients $c_k(\zeta)$ start at O$(\zeta^k)$ , whereas the remainder
(analytic) terms $r_k(\zeta)$ contain all powers of $\zeta$.
Let us concentrate on the non-analytic term and, using (\ref{Ak}),
rearrange the power series as
\begin{equation}
\sum_{k=0}^\infty A_kc_k(\zeta)=A_0\sum_{n=0}^\infty u_m\zeta^m.
\end{equation}
For the first few coefficients we find
\begin{equation}
u_0=1\qquad u_1=p\qquad u_2=\frac{p^2}{4}\qquad
u_3=-\frac{p^2}{18}+\frac{p^3}{36}\,.
\end{equation}
Let us compare this to the first few coefficients of the original power series
\begin{equation}
s_0=1\qquad s_1=-p\qquad s_2=\frac{p^2}{4}\qquad
s_3=-\frac{p^2}{18}-\frac{p^3}{36}\,.
\end{equation}
We see that they agree up to a sign change.
This is not accidental. First of all we can see that the original differential
operator of the confluent Heun equation ${\cal D}X=0$ is
\begin{equation}
{\cal D}=z(z-1)\frac{{\rm d}^2}{{\rm d}z^2}+(2z-1)\frac{{\rm d}}{{\rm d}z}
+p(2z-1),
\end{equation}
which in the $\zeta$ variable becomes
\begin{equation}
{\cal D}=\zeta(\zeta-1)\frac{{\rm d}^2}{{\rm d}\zeta^2}
+(2\zeta-1)\frac{{\rm d}}{{\rm d}\zeta}
-p(2\zeta-1)
\end{equation}
i.e. it is the same operator with $p\to-p$.
Let us write the ansatz
\begin{equation}
X=-c(\zeta)\ln\zeta+r(\zeta),
\end{equation}
where we assume that the coefficient funcions $c(\zeta)$ and $r(\zeta)$ are analytic at $\zeta=0$ (power series). Applying the differential operator to this ansatz we get
\begin{equation}
{\cal D}c=0
\label{Dc}
\end{equation}
and
\begin{equation}
{\cal D}r=2(\zeta-1)\frac{{\rm d}c}{{\rm d}\zeta}+c\,.
\end{equation}
From (\ref{Dc}) we then see that $c(\zeta)$ satisfies the $-p$ Heun equation
and we find that
\begin{equation}
c(\zeta)=A_0{\rm HeunC}\big[-p,-2p,1,1,0,\zeta\big].
\end{equation}
This is an alternative way to see that regularity at $z=1$ requires $A_0=0$.

\subsection{Half-waves}

Of course, the perfect half-waves (HW) are sinusoidal. For $X(U)=\sin U$ the
nodes are at
\begin{equation}
U_1=0,\qquad U_2=\pi,\qquad U_3=2\pi\dots
\end{equation}
We know that
\begin{equation}
{\rm between\; }U_1\ {\rm and\ }U_2:\quad X(U)>0,\qquad X^{\prime\prime}(U)= -\sin U<0\,.
\end{equation}
This means that in this interval the function is {\it concave} (it bends
downwards). In this section we will use {\it convex} in the sense of the opposite
of concave: meaning that the function is bending upwards.
Then we see that
\begin{equation}
{\rm between\; }U_2\ {\rm and\ }U_3:\quad X(U)<0,\qquad X^{\prime\prime}(U)= -\sin U>0\,.
\end{equation}
In this interval the sine function is convex. Then the whole picture is
repeated periodically.

\subsubsection{\PT}

In the PT case the function satisfies the differential equation
\begin{equation}
\frac{{\rm d}^2}{{\rm d}U^2}X(U)=-\frac{m(m+1)}{\cosh^2 U}X(U)
\end{equation}
and we have the familiar solution
$ 
X(U)=P_m(\tanh U).
$ 
Let us denote the roots of the Legendre polinomial (for fixed $m$) by
$ 
y_1,\,y_2,\,\dots y_m\,.
$ 
Then the nodes of the PT solution are at
\begin{equation}
U_r={\rm arctanh}\, y_r,\qquad r=1,2,\dots m\,.
\end{equation}

With our normalization $X(-\infty)=1$ and so for $U<U_1$ $X(U)>0$, then:
\begin{equation}
U_1<U<U_2:\qquad X(U)<0,\quad X^{\prime\prime}(U)>0\,.
\end{equation}
Here the solution is convex, bends back upwards until $U_2$. From there on:
\begin{equation}
U_2<U<U_3:\qquad X(U)>0,\quad X^{\prime\prime}(U)<0\,.
\end{equation}
The solution is concave, bends back downwards until $U_3$. This alternating
pattern repeats itself until $U_m$. The solution in each of the intervals is
monotonic until the (in this interval unique) local minimum/maximum is reached;
then it is again monotonic towards the next node. This looks like a distorted
HW.

The main reason of the HW-behaviour is that $X(U)$ and $X^{\prime\prime}(U)$
have opposite sign.

Between $U_1$ and $U_m$ there are \emph{exactly} $m-1$ such distorted HWs. It is easy
to see that this $m-1$ counting is correct by plotting
\begin{equation}
m=1\qquad X(U)=-P_1(\tanh U)=-\tanh U\,.
\end{equation}
This solution is monotonic, there is obviously no HW as seen in FIG.\#7 of \cite{DM-1}.
Similarly one can plot
\begin{equation}
m=2\qquad X(U)=-\frac{1}{2}+\frac{3}{2}\tanh^2 U.
\end{equation}
There is a single HW between $U_1=-{\rm arctanh}(1/\sqrt{3})$ and
$U_2={\rm arctanh}(1/\sqrt{3})$, as seen in FIG.\#8 of \cite{DM-1}\,.

\smallskip
Now we turn to flyby, which our main interest in this paper.

\subsubsection{Derived \PT}

Here the solution satisfies\footnote{We take here $X(U)=X^1(U)$. The
considerations for the $X^2(U)$ coordinate are analogous.} 
\begin{equation}
\frac{{\rm d}^2}{{\rm d}U^2}X(U)=-k\tanh U\frac{1}{\cosh^2 U}X(U)\,.
\end{equation}
As we have seen, for the HW behaviour it is crucial that $X$ and
$X^{\prime\prime}$ have opposite sign. In this case this is satisfied only for
$U>0$. However, it turns out that this is not a problem.
In our normalization $U(-\infty)=1$ and so the solution starts off positive.
But since initially $X^{\prime\prime}$ is also positive, the solution curve is
{\it convex}, so it bends upwards and does not vanish for negative $U$.
After crossing $U=0$ the curve becomes {\it concave} and now bends downwards
until it reaches the first node $U_1$. All nodes of $X^1$ are positive~!
This means that all our qualitative considerations for the PT case are
valid also here: there are exactly $(m-1)$ HWs between $U_1$ and $U_m$.
 
%

\section{Higher-derivative profiles}\label{d>2Sec}

Similar results can be obtained for the 
higher-derivative profiles proposed in \cite{GibbHaw71,LongMemory,ZZHflyby}.

$\bullet$ The system proposed by Braginsky and Thorne \cite{BraTho}  corresponds  \cite{GibbHaw71} to the \emph{2nd derivative of the Gaussian},
\beq
{\cal A}^{BT}(U)= 
\frac{\;\;d^2}{dU^2}\left(
 \frac{k}{\sqrt{\pi}}\,e^{-U^2}\right)\,.
\label{d=2BT}
\eeq

A surprising feature which extends our observation made before in the underived case and illustrated in FIG.\#(20) of \cite{DM-1} is that DM requires to putting one or the other component  to identically zero: and we get \emph{half-DM}, as shown in FIG.\ref{d2-Gausstraj}.
 
\begin{figure}[h]
\includegraphics[scale=.34]{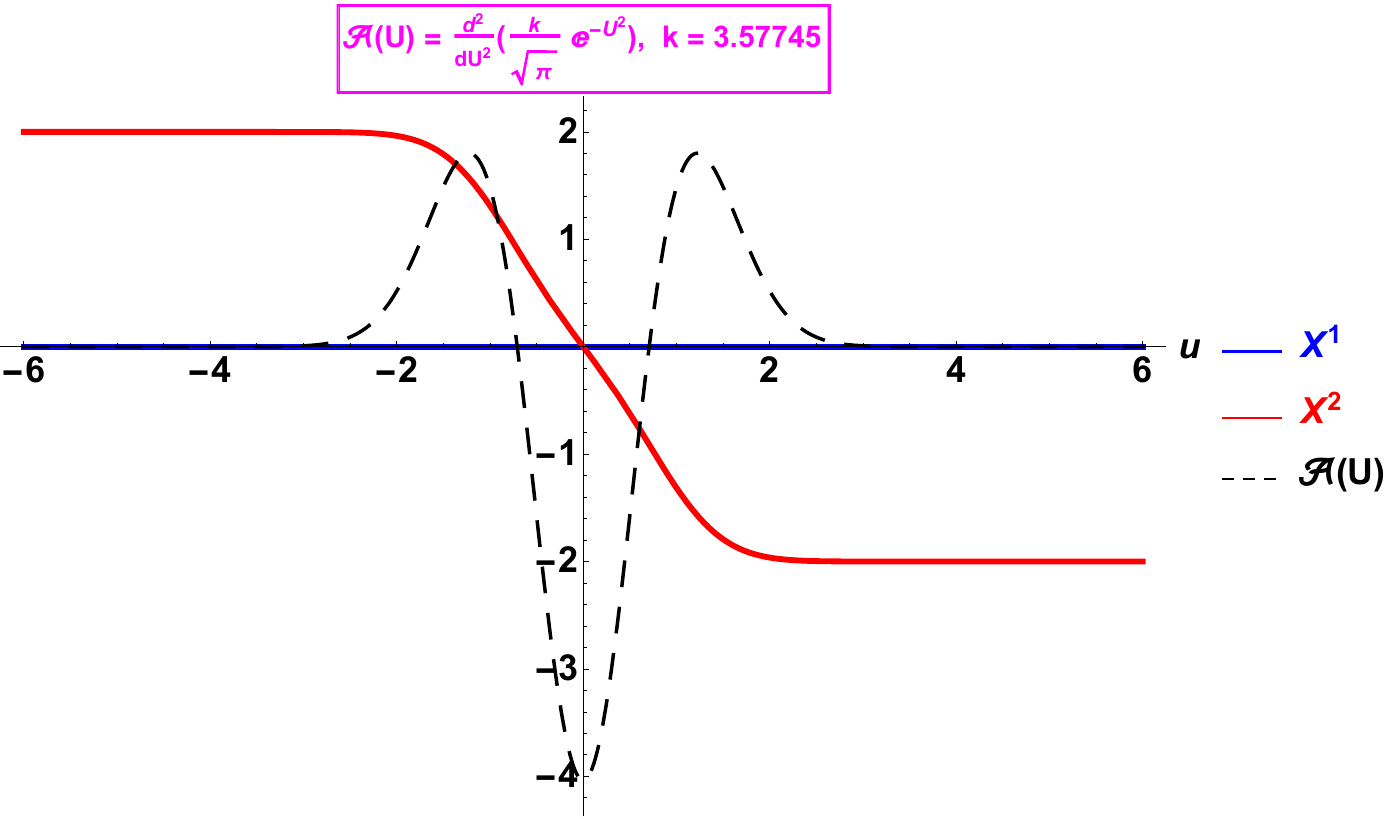}
\includegraphics[scale=.34]{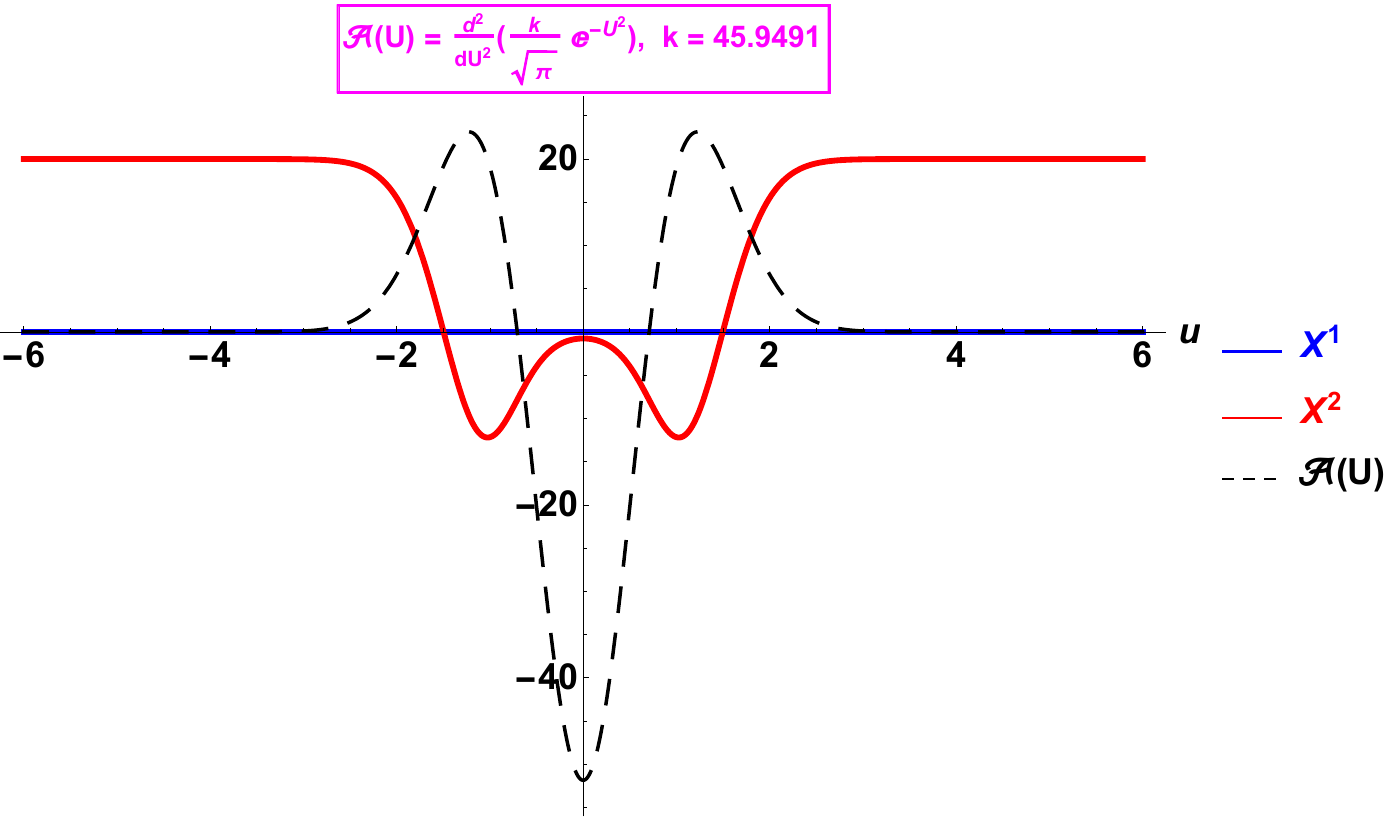}
\vskip-3mm
\hskip-15.2mm \magenta{\bf m = 1}\hskip 72mm
\magenta{\bf m=2}
\vskip-4mm
\caption{\textit{\small  The 2nd-derivative profile \eqref{d=2BT} proposed by Gibbons and Hawking \cite{GibbHaw71} for the Braginsky and Thorne theory \cite{BraTho} has even profile.  
We can get DM for one (say \red{${\bf X^2}$}) coordinate but then the other one should be put to zero, \blue{${\bf X^1}\equiv 0$}.}
\label{d2-Gausstraj}
}
\end{figure}

This curious ``half-sidedness'' is explained as follows. Assuming that $k >0$, DM is obtained for $X^2$ by fine-tuning the amplitude, $k=k_{crit}^{(2)}$. This does not work for $X^1$, though, unless $X^1\equiv0$, as confirmed by Figs.\ref{d2-Gausstraj} and FIG.\ref{2good1bad}.
\begin{figure}[h]
\includegraphics[scale=.32]{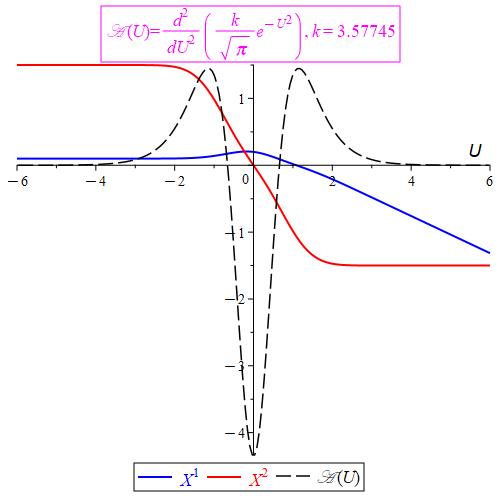} 
\vskip-3mm\caption{\textit{\small Fine-tuning one of the sectors for DM yields VM for the other, when the latter trajectory is identically zero.
}
\label{2good1bad} 
}
\end{figure}
However we could proceed also with the opposite casting and fine-tune the amplitude for $X^1$ to get $k=k_{crit}^{(1)}$. But the profiles have  opposite signs, therefore
\beq
k_{crit}^{(1)} = -k_{crit}^{(2)}\,.
\label{evenk}
\eeq
 However we have just one $k$, and the two choices are consistent only for $k=0$, -- \ie, for no wave at all.
The only way out is to turn off one of the components, either $X^1\equiv0$, 
or $X^2\equiv0$, depending on the sign of $k$~: 
 we get \emph{half-DM}. Taking $k < 0$ the sectors are interchanged.

\medskip
Not surprisingly, things work along the same lines for the \PT counterpart,
\beq
 \frac{\;d^2\cA^{PT}}{dU^2}= k\Big[(1-\tanh^2U)(1-3\tanh^2U)\Big]\,,
\eeq
for which the coefficient of the linear-in-$X$ term in \eqref{dPTeq} is replaced by   
 $1-3t^2$  cf. \eqref{A1PT}, which, for $t$ close to $\pm1$, multiplies the amplitude by roughly $(-2)$.
The trajectories shown in  FIG.\ref{2PT-m1m2}  are  similar to  their Gaussian-based counterparts in FIG.\ref{d2-Gausstraj}; the only difference is their slightly higher critical amplitude.
Numerical evidence suggests that  DM trajectories are composed again of (approximately) integer numbers of half-waves.

\begin{figure}[h]
\begin{center}
\end{center}
\includegraphics[scale=.28]{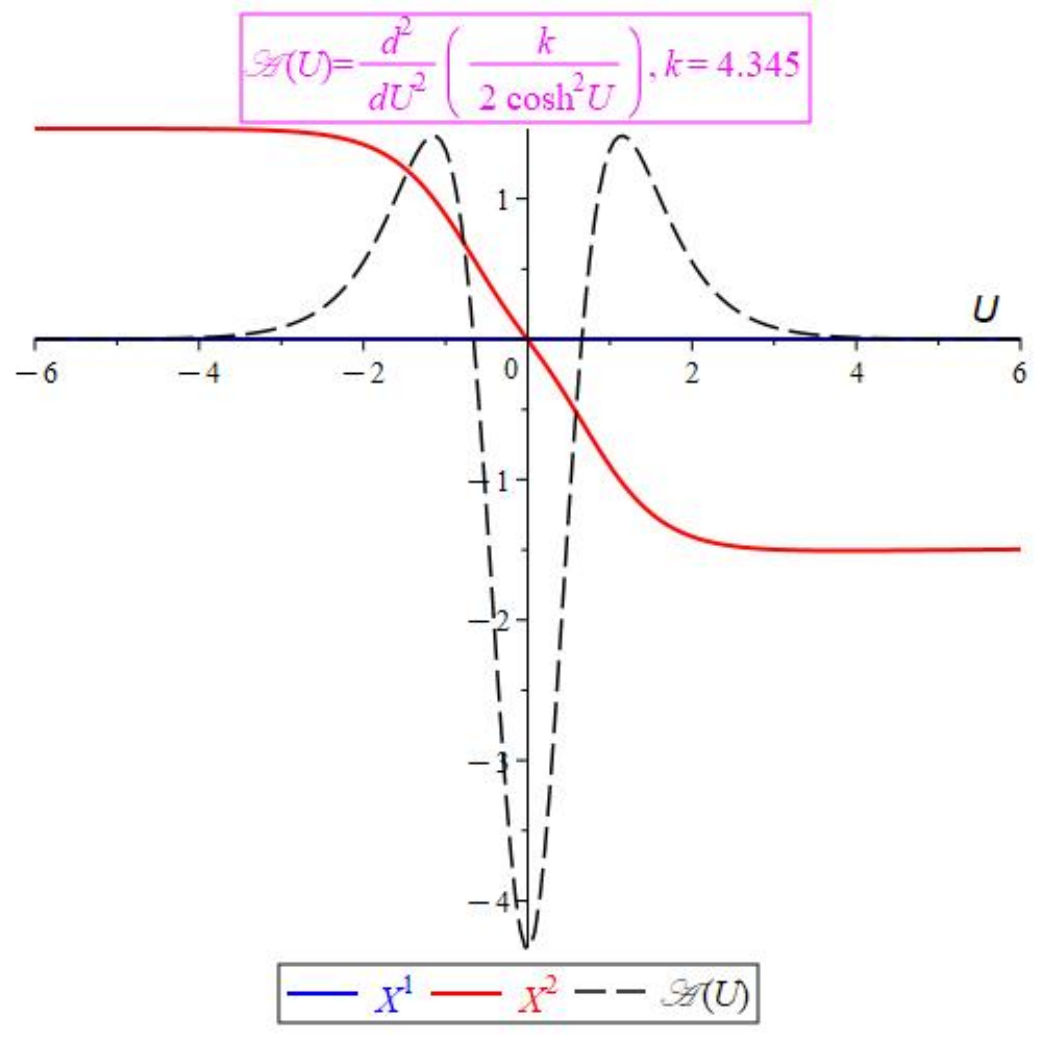}\qquad
\includegraphics[scale=.28]{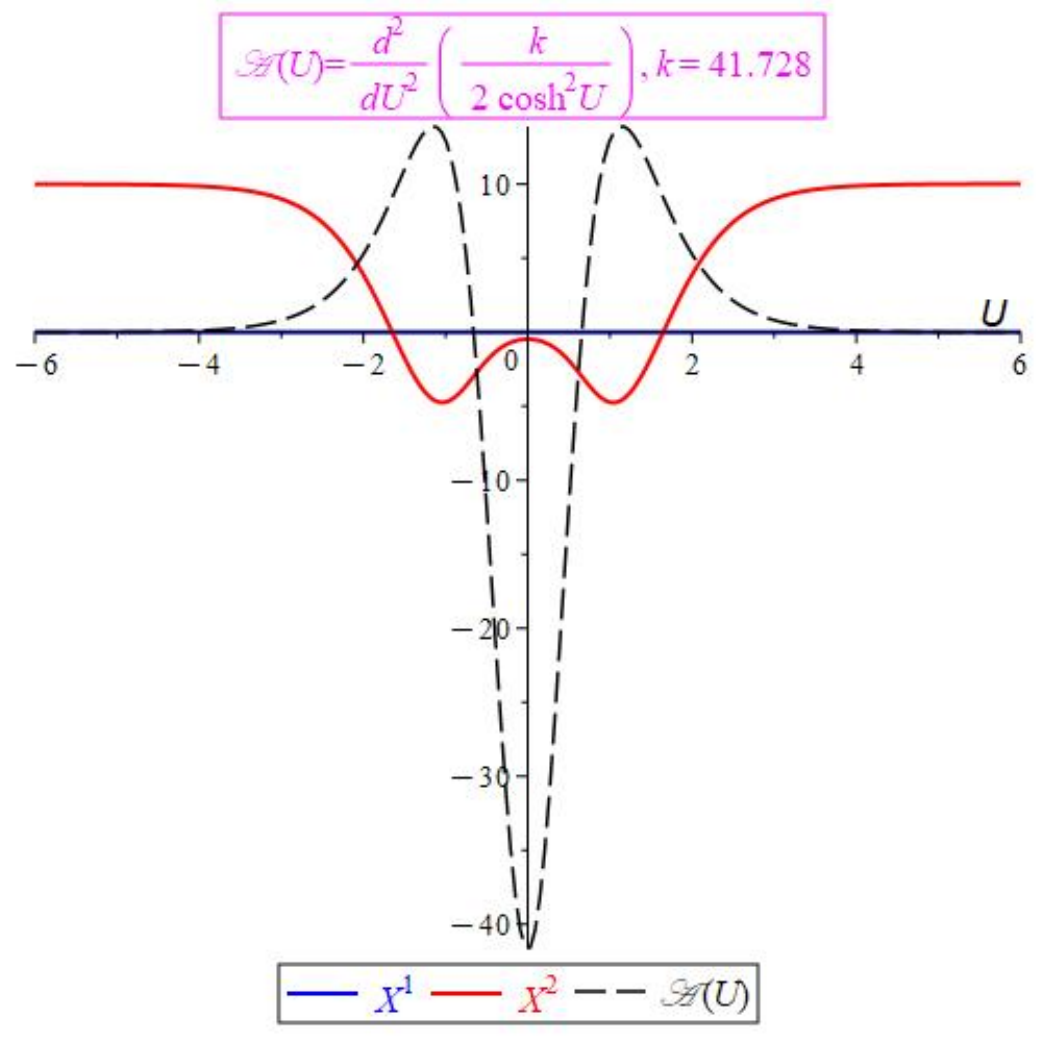}
{}\vskip-3mm\hskip-1mm
\magenta{\bf m = 1}\hskip47mm
\magenta{\bf m=2}
\vskip-3mm
\caption{\textit{\small The trajectories for the 2nd derivative of \PT  profile are similar to the Gaussian-based versions proposed by 
 Braginsky and Thorne \cite{BraTho,GibbHaw71}, shown in  FIG.\ref{d2-Gausstraj}.
} 
\label{2PT-m1m2}
}
\end{figure}

\medskip
$\bullet$ The \emph{3rd derivative of the Gaussian} 
\beq
{\cA}^{GC}(U) =
\frac{\;\;\,d^3}{dU^3}\!\left(
 \frac{k}{\sqrt{\pi}}\,e^{-U^2}\right) 
\label{Kd3}
\eeq
proposed to describe \emph{gravitational collapse} \cite{GibbHaw71}
was studied in some detail in \cite{ShortMemory, LongMemory} exhibits \emph{DM} again for both components, as shown in
FIG.\ref{d3x1x2}. 
 The plot of its \PT counterpart is very similar up to having somewhat higher critical parameters, 
$k_1=4.7457$ and $k_2=31.14958$, respectively. It is therefore omitted. 
\begin{figure}[h]
\includegraphics[scale=.345]{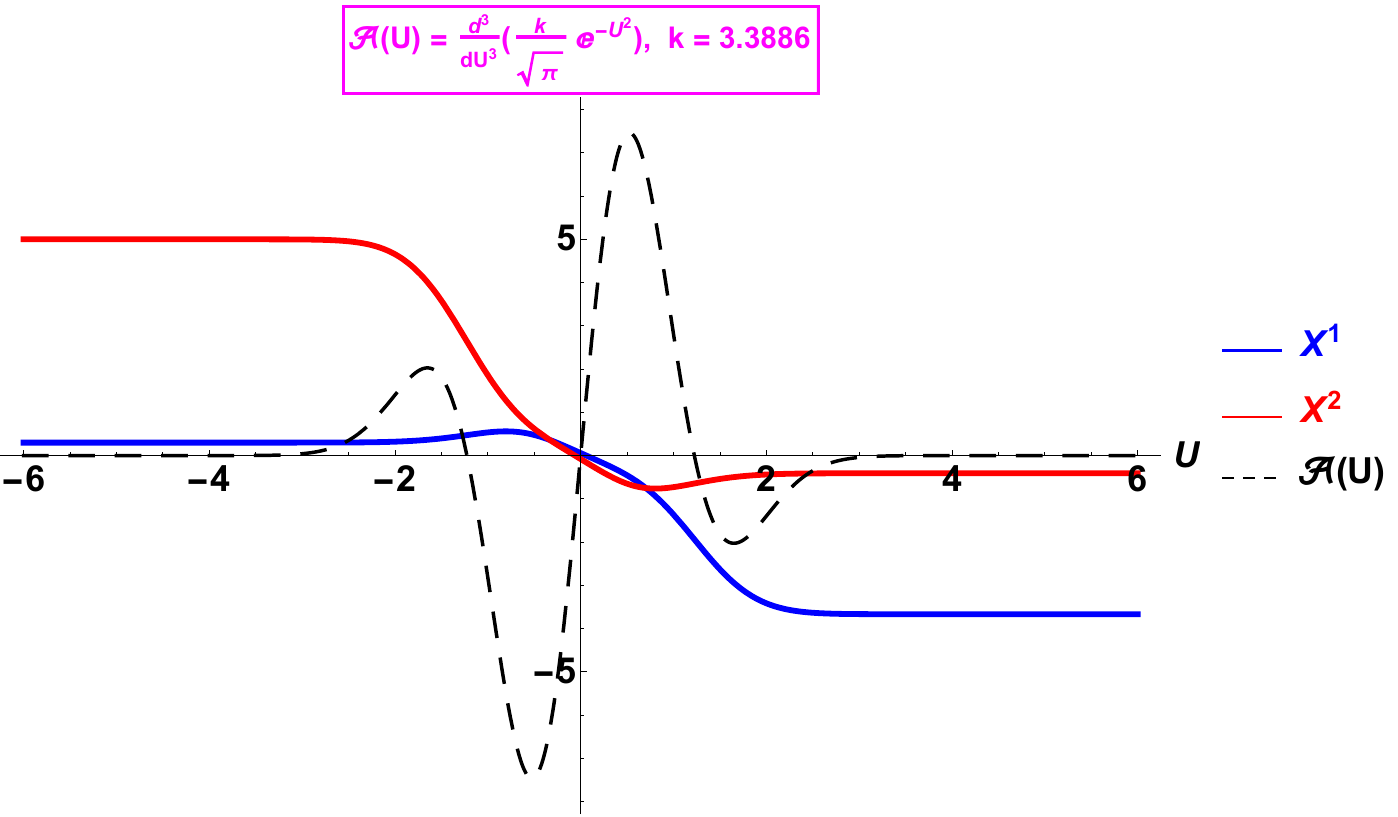}
\includegraphics[scale=.345]{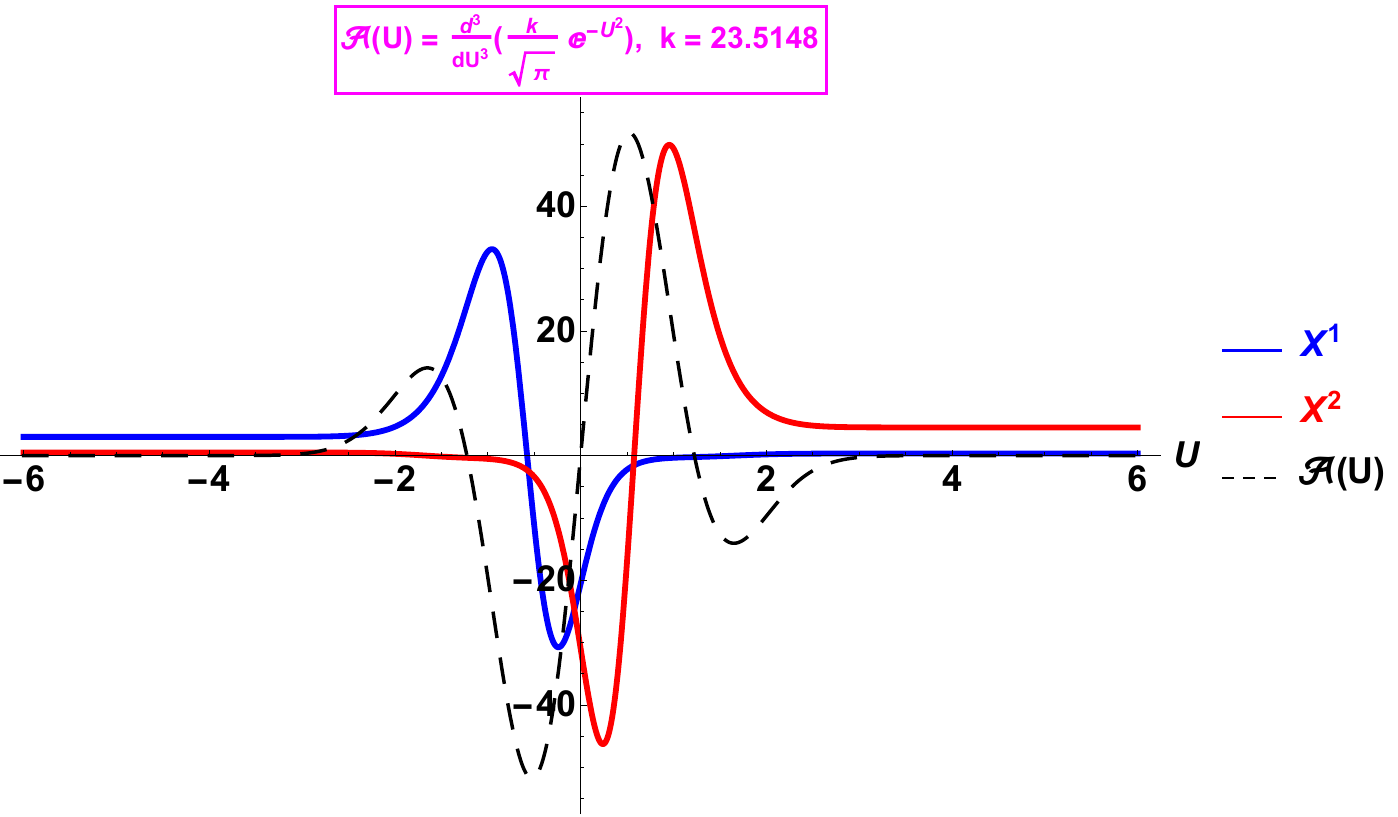}
{}\vskip-3mm
\hskip-15mm \magenta{\bf m = 1}\hskip 72mm
\magenta{\bf m=2}
\vskip-4mm
\caption{\textit{\small The 3rd derivative of the Gaussian was proposed for gravitational collapse  \cite{GibbHaw71}. For  $k_{crit}$  the velocity vanishes at $U=\pm\infty$, as required for DM in both components, \eqref{DMcond}.} 
\label{d3x1x2}
}
\end{figure}
\goodbreak

Models with higher-order derivatives of the Gaussian  follow  a similar pattern. For the 4th order derivative profile, for example, we would get again  {half-DM effect}, etc. We now explain the reason by studying the behavior under reversing the sign of $U$, 
\beq
U \to - U\,.
\label{Urev}
\eeq

We start with an even seed profile, 
\beq
{\cA}^0(-U)= {\cA}^0(U)\,  
\label{evenseed}
\eeq
which could be either the  Gaussian or  \PT\! as in \cite{DM-1}, for example.
Then we consider the profile obtained by deriving $\cA^0$
$\ell$ times,
\begin{equation}
\cA^{\ell}=
\frac{\;\;\,d^{\ell}\!\cA^0}{\; dU^{\ell}}\,.
\label{dellA}
\end{equation}
The sign-reversal \eqref{Urev} leaves invariant the 2nd-order $U$-derivative terms in \eqref{Bgeoeqn} but has an $\ell$-dependent effect on the equations of motion~:

$\bullet$ for $\ell=2n+1,\,n=0,\,1, \dots$ \emph{odd}, the profile is \emph{antisymmetric},
\beq
{\cA}^{2n+1}(-U)= -{\cA}^{2n+1}(U), 
\label{cAodd}
\eeq
implying that the two transverse components in \eqref{geoX1}-\eqref{geoX2} are merely interchanged under \eqref{Urev} (as noticed for flyby in sect.\ref{dPTsec}). Thus they satisfy the same pair of equations up to permutation. Therefore the DM condition holds either for both or for none of them. When it does, then  ``\emph{full DM}'' is obtained for both coordinates
  as illustrated in FIGs. \ref{d1-PT-m12}-\ref{dPT166-276} for $\ell=1$, and in FIG.\ref{d3x1x2} for $\ell=3$. 

$U$-reversal symmetry \eqref{Urev} allows us to generalize an observation made in sect.\ref{analSec}: for the chosen normalization the $U \to -U$ symmetry implies 
\beq
X^1(\infty)\,X^2(\infty)=1\,.
\label{recipr12}
\eeq
Thus the larger one of the components is, the smaller the other will be. A higher amplitude improves the precision of DM.    

The parity-dependent behavior is in fact quite general, as it could also be illustrated for example, by the sequence \eqref{dellA} with odd seed profile,
$
\cA^0 \propto 
\tanh U\, 
$ 
\cite{Carroll4GW}.  

$\bullet$ for  $\ell=2n,\, n=0,\,1, \dots$, \emph{even}, the profile is \emph{symmetric}, 
\beq
{\cA}^{2n}(-U)= {\cA}^{2n}(U)\,.  
\label{cAeven}
\eeq
Each of the transverse equations \eqref{geoX1} and \eqref{geoX2} is left invariant under \eqref{Urev} 
  
DM can be obtained by fine-tuning the amplitude in one of the sectors but then the other should be put to zero: we get ``\emph{half-DM}'', as illustrated   in FIG.\# 20 of \cite{DM-1} for $\ell=0$ and in FIG.\ref{d2-Gausstraj} for $\ell=2$.
\goodbreak

\section{Longitudinal motion}\label{Vmotion}
 
 So far we studied motion  only in the transverse plane.
  What about the ``vertical'' coordinate ? We summarize the answer given in \cite{DM-1}.  
The projection of motion to transverse space  is independent of $V$ and the  parallel-transport eqn. \eqref{geoV} implies, for $\fm=0$, that
the vertical motion is given by, 
\beq
\hat{V}(U)=
 V_0 - \int_{-\infty}^U\!\!{\cL}_{NR}\,du\,,
\label{nullV}
\eeq
where ${\cL}_{NR}$ is the underlying non-relativistic  Lagrangian, see \# (V.1) of \cite{DM-1}.
We proved also that the DM conditions 
$
X^{\prime}(U=-\infty)=0=X^{\prime}(U=\infty)\,
$ in \eqref{DMcond}
imply that for $U$ in the Afterzone the integral in 
 \eqref{nullV} vanishes~: 
the classical Hamiltonian action calculated along the underlying NR trajectory should vanish,
\beq
\displaystyle{\int_{-\infty}^{\;U}}\!\cL_{NR}\,du=0\,,
\for U\geq U_a
\label{noaction}
\eeq 
 leaving us with
$\hat{V}(U)=V_0$. Thus for a massless particle {the vertical light-cone coordinate $V$ does not move} at all.

Turning to massive geodesics  characterized by the Jacobi invariant  $\fm>0$ in \eqref{Jacobiinv}, the $V$-motion is, see \# (VI.5) in \cite{DM-1}, a mere linear 
shifted,
\beq
V(U) = V_{0} - \frac{1}{2}\left(\frac{\fm}{M}\right)^{2}U \,,
\label{massiveV}
\eeq
where $M$ is the conserved quantity generated by the Killing vector $\p_V$, interpreted, in the ``Bargmann'' framework \cite{DBKP,DGH91}, as the underlying NR mass.
The massive trajectories are ``tilted'', as shown in FIG.\#(13) of \cite{DM-1}. However
when the relativistic and non-relativistic masses are  identical,   
\beq
\fm = M\,,
\label{fmM}
\eeq
then, setting 
\beq
Z=V +\half U
\label{ZVhalf}
\eeq
we get,
\beq
Z(U) =  V_0 = \const
\label{Zmfix}
\eeq 
\ie, we get DM also in the longitudinal $Z$-direction,
see FIG.\#(14) of \cite{DM-1}.

\section{Conclusion}\label{Concl}

Test particles hit by a sandwich  \GW exhibit, generically, the velocity effect (VM) \cite{Ehlers,Sou73,AiBalasin}~: after the wave has passed, the particles fly apart with diverging constant, non-zero velocity \cite{ShortMemory,LongMemory}. Zel'dovich and Polnarev suggested instead  that flyby would approximately generate pure displacement ({DM}) with vanishing relative velocity \cite{ZelPol}. 
Our main result is to confirm and indeed refine the statement of Zel'dovich and Polnarev.
Both numerical and analytical evidence show that a \emph{judicious choice of the wave parameters} 
 yields pure displacement \cite{DM-1} 
 
We focus our  attention 
 at linearly polarized \GWs whose profile is a derivative of order $\ell=0,\, 1,\ \dots$ of a seed profile $\cA^{0}$, \eqref{dellA}. The latter can be, for example, a Gaussian  \cite{GibbHaw71} or its analytical Doppelg\"anger, the \PT potential \cite{PTeller,Chakra,DM-1,ZZHflyby}.  Other promising candidates (not studied in the paper) could be Rosen-Morse type \cite{Scarf} or inverse-power \cite{AndrPrenc,Ilderton,ZhaoU-4} profiles. Analytically solvable models can be built also from square-profile \cite{Chakraborty:2022qvv,ZZHflyby}.
 
This paper extends the above results to a more physical context.
We find that DM solutions are composed approximately of an integer number of half-waves which is exact for the simple \PT profile \eqref{APT} and holds approximately for $\ell =1$, as said in \cite{ZelPol}.
Particular attention is paid to flyby \cite{ZelPol}, which might approach observability  \cite{Favata08,Favata10,Lasenby,Lasky,LISAflyby}.

In the generic {VM} case the particles pick up some energy, which might lead to the absorption of the wave, as discussed quite some time ago. Sect. 2-5.8 of the seminal review of Ehlers and Kundt \cite{Ehlers} anticipates  much of later work  \footnote{We are grateful to P.C. Aichelburg for calling our attention at this.}. Although they did not identify either the Carroll symmetry \cite{Leblond,SenGupta,BacryLL,Carrollvs,Carroll4GW} or {DM}, their intuitive argument in their subsec.{2-5.8} point in this direction:

\begin{quote} \textit{\narrower 
\dots one finds the effect of a wave pulse on a cloud of resting particles: [their  eqn. (2-5.58)] gives the amount of \underline{transverse momentum, longitudinal momentum}, and \underline{kinetic energy} [\dots] a particle has gained [\dots] 
after the passage of the pulse. This consideration suggests convincingly that a cloud of dust is \underline{able to extract energy from a gravitational wave} \cite{Ehlers}.
}
\end{quote} 

 Preliminary versions of this research \cite{ZZHflyby} were presented by PAH at the Wigner Institute in Budapest (April 5 2024), at the Workshop
{\sl Carrollian Physics and Holography}${}_{-}$CDFG${}_{-}$2024 organised at the Schr\"odinger Institute in Vienna (April 17  2024), and at the conference {\sl ``Conformal anomalies: theory and applications 2024,
https://indico.math.cnrs.fr/event/10718/} Tours (May 7 2024).  
 P-M Zhang presented related results 
in \cite{ZZHflyby}.

Related ideas were put forward  also in \cite{Bondi57,Sou73,AiBalasin,GriPol}.
Recent results on the Memory Effect and the DM or VM puzzle include refs. \cite{Chakraborty:2022qvv,DM-1,Jibril,ZZHflyby, Mitman,HarteOancea,Arpan,DeyKar}.

\goodbreak
 
\vskip-6mm
\begin{acknowledgments}\vskip-4mm
 We are indebted to  G.~Gibbons for his insights and advices. We benefitted from discussions and correspondence  also with P.~C. Aichelburg, T. Damour, L. Di\'osi, M. Elbistan, S. Kar, B. Kocsis, V. Komornik, K. Mitman and S. Silagadze. JB is grateful to SYSU for hospitality extended to him in Zhuhai. 
 PAH thanks the Erwin Schr\"odinger Institute (ESI, Vienna) for hospitality during the Workshop
 {\sl Carrollian Physics and Holography}${}_{-}$CDFG${}_{-}$2024. 

\end{acknowledgments}
\goodbreak



\end{document}